\documentclass[prl,aps,twocolumn,preprintnumbers, showpacs, nofootinbib,superscriptaddress,notitlepage]{revtex4-1}
\usepackage{amssymb, amsthm}
\usepackage{amsmath}    
\usepackage{mathrsfs}   
\usepackage{graphicx}   
\usepackage{color}      
\usepackage{colortbl}
\usepackage{footnote}   
\usepackage{slashed}    

\usepackage[utf8]{inputenc}
\usepackage[dvipsnames]{xcolor}
\usepackage[normalem]{ulem}

\usepackage{subfiles}

\begin{document}


\title{Unpolarized gluon PDF of the nucleon from lattice QCD  at physical point in the continuum limit}

\author{\includegraphics[scale=0.1]
{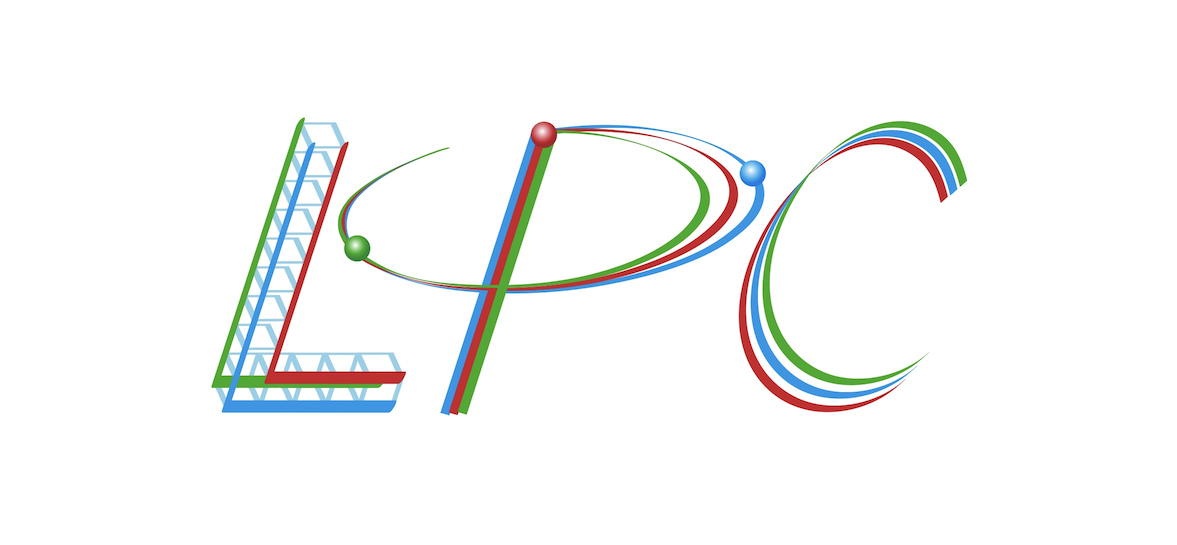}\\Chen Chen}
\affiliation{Institute of Modern Physics, Chinese Academy of Sciences, Lanzhou 730000, China}
\affiliation{University of Chinese Academy of Sciences, School of Nuclear Science and Technology, Beijing 100049, China}

\author{Chunhua Zeng}
\affiliation{Institute of Modern Physics, Chinese Academy of Sciences, Lanzhou 730000, China}
\affiliation{University of Chinese Academy of Sciences, School of Nuclear Science and Technology, Beijing 100049, China}

\author{Hongxin Dong}
\affiliation{Institute of Modern Physics, Chinese Academy of Sciences, Lanzhou 730000, China}
\affiliation{Nanjing Normal University, Nanjing, Jiangsu 210023, China}

\author{Liuming Liu}
\email[Corresponding author: ]{liuming@impcas.ac.cn}
\affiliation{Institute of Modern Physics, Chinese Academy of Sciences, Lanzhou 730000, China}
\affiliation{University of Chinese Academy of Sciences, School of Nuclear Science and Technology, Beijing 100049, China}
\author{Xiaomin Shen}
\affiliation{Institute of Modern Physics, Chinese Academy of Sciences, Lanzhou 730000, China}
\affiliation{University of Chinese Academy of Sciences, School of Nuclear Science and Technology, Beijing 100049, China}
\author{Peng Sun}
\email[Corresponding author: ]{pengsun@impcas.ac.cn}
\affiliation{Institute of Modern Physics, Chinese Academy of Sciences, Lanzhou 730000, China}
\affiliation{University of Chinese Academy of Sciences, School of Nuclear Science and Technology, Beijing 100049, China}
\author{Xiaonu Xiong}
\affiliation{School of Physics and Electronics, Central South University, Changsha 418003, China}
\author{Yi-Bo Yang}
\email[Corresponding author: ]{ybyang@itp.ac.cn}
\affiliation{University of Chinese Academy of Sciences, School of Physical Sciences, Beijing 100049, China}
\affiliation{CAS Key Laboratory of Theoretical Physics, Institute of Theoretical Physics, Chinese Academy of Sciences, Beijing 100190, China}
\affiliation{School of Fundamental Physics and Mathematical Sciences, Hangzhou Institute for Advanced Study, UCAS, Hangzhou 310024, China}
\affiliation{International Centre for Theoretical Physics Asia-Pacific, Beijing/Hangzhou, China}

\author{Fei Yao}
\affiliation{Physics Department, Brookhaven National Laboratory, Upton, New York 11973, USA}

\author{Jian-Hui Zhang}
\email[Corresponding author: ]
{zhangjianhui@cuhk.edu.cn}
\affiliation{School of Science and Engineering, The Chinese University of Hong Kong, Shenzhen 518172, China}

\author{Shiyi Zhong}
\affiliation{School of Science and Engineering, The Chinese University of Hong Kong, Shenzhen 518172, China}

\begin{abstract}
We report a state-of-the-art lattice QCD calculation of the nucleon unpolarized gluon parton distribution function employing large-momentum effective theory. The calculation is carried out on the 2+1 flavor CLQCD ensembles with five lattice spacings $a=\{0.105,0.0897,0.0775, 0.0688, 0.0519\}$ fm and various pion masses ranging from $136$ MeV to $317$ MeV, covering nulceon momenta up to $3$ GeV. Distillation technique is applied to improve the signal of two-point correlators. 
We then apply the state-of-the-art hybrid renormalization and one-loop perturbative matching, and extrapolate the result to the continuum limit, infinite momentum limit and physical pion mass. 
\end{abstract}

\maketitle

{\em Introduction:}
Parton distribution functions (PDFs) are fundamental quantities that encode the internal structure of hadrons in terms of their constituent quarks and gluons. Among these, the gluon PDF plays a particularly crucial role in our understanding of quantum chromodynamics (QCD) and has profound implications for high-energy physics phenomenology. 
It represents the probability density of finding a gluon carrying a fraction $x$ of the nucleon's momentum, and provides essential input for theoretical predictions in hadron-hadron collisions, deep inelastic scattering processes, and precision tests of the Standard Model at hadron colliders.

Despite its fundamental importance, the determination of the gluon PDF has historically presented significant theoretical and computational challenges. Traditional approaches rely on global fits to experimental data from deep inelastic scattering and hadron-hadron collisions~\cite{Bailey:2020ooq, Hou:2019efy,NNPDF:2017mvq,NNPDF:2021njg,Accardi:2016qay,Dulat:2015mca}. However, these phenomenological extractions are inherently model-dependent and face severe limitations in the large-$x$ region where experimental data is sparse and the gluon PDF becomes increasingly uncertain. The lack of direct experimental constraints on the gluon PDF at large $x$ has motivated the development of first-principles theoretical approaches.

The advent of Large Momentum Effective Theory (LaMET)~\cite{Ji:2013dva, Ji:2014gla, Ji:2017rah} has provided a promising framework for direct first-principles calculations of PDFs from lattice QCD. It offers a systematic procedure to extract light-cone PDFs from quasi-PDFs, where the latter can be computed directly on the lattice and matched to the former through perturbative matching relations. The key insight of LaMET is that when a hadron carries large momentum $P_z \gg \Lambda_{\text{QCD}}$, the quasi-PDF approaches the light-cone PDF up to perturbative matching and power corrections that can be systematically controlled. Recent years have witnessed remarkable progress in lattice PDF calculations, with numerous studies successfully extracting quark PDFs for both nucleons and mesons~\cite{Lin:2018pvv,LatticeParton:2018gjr, LatticeParton:2020uhz, LatticePartonLPC:2021gpi, Hua:2020gnw, LatticePartonLPC:2022eev, Zhang:2022xuw, LatticeParton:2022zqc, LatticePartonLPC:2023pdv, LatticeParton:2022xsd, LatticePartonCollaborationLPC:2022myp, LatticeParton:2023xdl, LatticePartonLPC:2023pdv, Zhang:2024omt, Chen:2024rgi, LatticeParton:2024zko, LPC:2025spt, Zhang:2025hvf, Good:2024iur}. 

However, lattice calculations of gluon PDFs present unique challenges that have limited their progress compared to quark PDFs. Although the gluon quasi-PDF operators have been proposed and extensively studied~\cite{Wang:2017qyg,Wang:2017eel,Zhang:2018diq,Wang:2019tgg}, their matrix elements are notoriously noisy in lattice simulations, requiring sophisticated techniques to extract clean signals. %
Several pioneering studies~\cite{Fan:2018dxu, Fan:2020cpa, HadStruc:2021wmh, Fan:2022kcb, Good:2023gai, Delmar:2023agv, Fan:2021bcr, Good:2023ecp, Salas-Chavira:2021wui} have made significant progress in gluon PDF calculations, with most employing the pseudo-PDF method~\cite{Radyushkin:2017cyf}. LaMET calculations of the gluon PDF have recently appeared~\cite{Good:2024iur, Good:2025daz, NieMiera:2025mwj}, which have been performed at unphysical  pion mass.

In this paper, we present a state-of-the-art lattice QCD calculation of the unpolarized gluon PDF of the nucleon using LaMET, where the continuum limit, infinite momentum limit and extrapolation to the physical pion mass have been taken. This is crucial for obtaining a light-cone PDF that can be compared with experimental extractions using global fits. We perform our calculation on ensembles with five lattice spacings ranging from $0.0519$ fm to $0.105$ fm  and pion masses  ranging from $136$ MeV to $317$ MeV. We use several nucleon momenta up to 3~GeV to perform the extrapolation to the infinite-momentum limit. The distillation quark smearing method~\cite{HadronSpectrum:2009krc} and momentum smearing technique~\cite{Egerer:2020hnc} are applied to improve the signal-to-noise ratio of the gluonic matrix elements. We employ a state-of-the-art hybrid renormalization scheme~\cite{Ji:2020brr} to remove ultraviolet divergences nonperturbatively, and apply one-loop perturbative matching to obtain the light-cone PDF. Our results represent a significant step forward in the precision determination of gluon PDFs from first principles and provide the most complete lattice QCD calculation of the nucleon gluon PDF to date.

{\em Lattice calculations:} In this work we use the 2 + 1 flavor QCD
ensembles with tadpole improved tree level Symanzik (TITLS) gauge action and the tadpole improved tree level Clover (TITLC) fermion action  generated by the CLQCD collaboration ~\cite{CLQCD:2023sdb,CLQCD:2024yyn}.  The calculation is performed on eight  ensembles with five lattice spacings ranging from $0.0519$ fm to $0.105$ fm  and pion masses ranging from  $136$ MeV to $317$ MeV.  The detailed information is given in table ~\ref{Tab:setup}.

\renewcommand{\arraystretch}{1.5}
\begin{table}
\footnotesize
\centering
\begin{tabular}{cclcccccc}

\hline
\hline
Ensemble ~&$a$(fm) ~& \ \!$L^3\times T$  ~& $m_\pi$(MeV) ~& $m_\pi L$ & $N_{\text{conf.}}$ \\  



\hline
C24P29  ~& $0.105$~& $24^3\times 72$  ~& $292.3(1.0)$   ~&$3.746$ & 878                \\  
\hline
C32P29  ~& $0.105$~& $32^3\times 64$  ~&  $292.1(0.8)$   ~& $4.991$  &   980              \\  

\hline
C32P23  ~& $0.105$~& $32^3\times 64$  ~& $227.9(1.2)$   ~&$3.894$ &   597              \\  

\hline
C48P14  ~& $0.105$~& $48^3\times 96$  ~& $136.4(1.7)$   ~&$3.495$ &  347   \\
\hline 

E32P29  ~& $0.0897$~& $32^3\times 64$  ~&  $287.3(2.5)$~&  $4.179$ & 1700         \\ 

\hline
F32P30  ~& 0.0775~& $32^3\times 96$  ~& $300.4(1.2)$  ~& $3.780$ & 777\\
\hline 

G36P29  ~& $0.0688$~& $36^3\times 108$  ~&  $297.2(0.9)$~&  $3.731$ & 349          \\ 

\hline
H48P32   ~& 0.0519~& $48^3\times 144$  ~& $316.6(1.0)$  ~& $4.000$ & 650
            \\  


\arrayrulecolor{gray!50}

\arrayrulecolor{black}
\hline

\end{tabular}
 \caption{The simulation setup, including lattice spacing $a$, lattice size $L^3\times T$ , $\pi$ mass $m_{\pi}$ and the number of configurations for each ensemble.}
 \label{Tab:setup}
\end{table}

According to LaMET, to obtain the unpolarized gluon PDF, one starts by calculating the bare quasi-light-front correlator
\begin{equation}
\tilde{h}_B(z,P_z,1/a)=\langle P_z|O(z)|P_z \rangle,
\end{equation} 
where the gluon operator $O(z)$ takes the following form which has been shown to be multiplicatively renormalizable~\cite{Zhang:2018diq,Balitsky:2019krf}
\begin{align}
O(z)=M_{tx;tx}(z)+M_{ty;ty}(z)-2M_{xy;xy}(z),
\label{ope_com}
\end{align}
with
\begin{align}
M_{\mu \lambda; \nu \rho}(z) &=\sum_{\vec{x}}Tr[F_{\mu\lambda}(\vec{x}+z\hat{z})
U(\vec{x}+z\hat{z},\vec{x}) \nonumber \\  & F_{\nu \rho}(\vec{x})U(\vec{x},\vec{x}+z\hat{z})], 
\label{M_fund}
\end{align}
for the fundamental representation. $U(\vec{x}+z\hat{z},\vec{x})$ is the Wilson line introduced to ensure gauge invariance and $F_{\mu\nu}=(\partial_{\mu}A_{\nu}^a-\partial_{\nu}A_{\mu}^a-gf^{abc}A_{\mu}^bA_{\nu}^c)T^a$ is the gauge field tensor. We employ the nucleon interpolating operator $N = P_+(u^TC\gamma_5 \gamma_t d)u$ , where $C$ denotes the charge-conjugation operator and $P_+ = \frac{1}{2}(1+\gamma_t)$ is the positive parity projector. As shown in Ref.~\cite{Zhang:2025hyo}, this operator has  kinematically-enhanced  overlap with the ground state at large momentum. The quark propagators are computed for all time slices using the distillation quark smearing method~\cite{HadronSpectrum:2009krc}. The number of eigenvectors $N_{ev}$ in the smearing operator is 200 for the ensembles C48P14, G36P29 and H48P32, and $N_{ev}=100$ for the remaining ensembles. 
To enhance the signal at higher momenta, we apply the momentum smearing technique within distillation, where the eigenvectors are multiplied by a  phase shift following Ref.\cite{Egerer:2020hnc},
\begin{align}
\tilde{\xi}_a^{(m)}(\vec{x},t)=e^{i\vec{\zeta}\cdot\vec{x}}\xi_a^{(m)}(\vec{x},t),
\end{align}
where $\xi_a^{(m)}$ is the $m$-th eigenvector with color index $a$ and $(\vec{x},t) $ represents space-time coordinate. $\vec{\zeta}=n_\mathrm{sm}\cdot\frac{2\pi}{L}\hat{z}$ is the smearing parameter. The detailed information of smearing parameter can be found in the supplemental material ~\cite{supplemental}.  Additionally,  ten-steps hypercubic (HYP) smearing is applied on the gluon operators in order to suppress potential errors from linear divergence. 

The bare matrix elements are then obtained by fitting the ratio of the three-point to two-point correlation functions using one-state Feynman-Hellmann method~\cite{Bouchard:2016heu}, the details of the fitting are provided in the supplemental material ~\cite{supplemental}.


{\em Renormalization:}
The bare matrix elements contain both linear and logarithmic divergences that must be removed via an appropriate renormalization procedure. 
Here we adopt the hybrid scheme proposed in Ref.~\cite{Ji:2020brr}. In this approach, the divergences at short distances are removed by dividing by the zero-momentum bare matrix element, while at long distances they are removed by a self-renormalization procedure~\cite{LatticePartonLPC:2021gpi}. The renormalized matrix element is thus defined as:

\begin{align}
\tilde{h}_R(z,P_z)
=&\frac{\tilde{h}^n_B(z,P_z,1/a)}{\tilde{h}^n_B(z,P_z=0,1/a)}\theta(z_s-|z|)\notag\\
&+\eta_s\frac{\tilde{h}^n_B(z,P_z,1/a)}{Z_R(z,1/a)} \theta(|z|-z_s), 
\label{eq:hybrid_scheme}
\end{align}
where $\tilde{h}^n_B(z,P_z,1/a)$ is the normalized bare matrix element,  $z_s$ is introduced to distinguish the short and long distances, and $\eta_s$ is  the scheme conversion factor that ensures the continuity of the renormalized quasi-LF correlation at $z=z_s$.
The self-renormalization factor $Z_R(z,1/a)$ 
takes the following parametrization form, which contains both the linear divergence and logarithmic UV divergences:
\begin{align}
\label{rnmlfctr}
Z_R & (z,1/a,\mu;k,\Lambda_{\rm QCD},m_0,d)=\exp\bigg\{\frac{kz}{a\ln[a\Lambda_{\text{QCD}}]}+m_0 z \notag\\
& +\frac{5C_A}{3b_0}\text{ln}\bigg[\frac{\text{ln}(1/a\Lambda_{\text{QCD}})}{\text{ln}(\mu/\Lambda_{\text{QCD}})}\bigg]+\frac{1}{2}\text{ln}\bigg[\bigg(1+\frac{d}{\text{ln}[a\Lambda_{\text{QCD}}]}\bigg)^2\bigg]\bigg\},
\end{align} 
where the first term is the linear divergence,  $m_0z$ denotes a finite-mass contribution from
the renormalization ambiguity, and the last two terms come from the
resummation of the leading and subleading logarithmic divergences. According to Ref.~\cite{LatticePartonLPC:2021gpi}, $k$ is related to the specific
discretized action, while $d$ and $m_0$ are determined by matching to the continuum scheme at short distances. 
In the $\overline{\text{MS}}$ scheme, the short-distance result takes the following form at next-to-leading order (NLO):
\begin{align}
\label{Zms}
\tilde{h}^{\rm pert}_{\overline{{\text{MS}}}}(z,\mu)=1+\frac{\alpha_s(\mu) C_A}{4\pi}\left(\frac{5}{3}\ln\frac{z^2\mu^2}{4e^{-2\gamma_E}}+3\right),
\end{align}
where $\gamma_E$ is the Euler-Mascheroni constant. We set the renormalization scale $\mu=2$~GeV, yielding a running coupling constant $\alpha_s(\mu)= 0.296$.

The parameters in $Z_R$ are determined by fitting the bare matrix elements  of the ensembles C24P29, E32P29, F32P30, G36P29 and H48P32 at zero momentum: 
\begin{align}
&\text{ln}\tilde{h}_B^n(z,P_z=0,1/a) 
 =  \text{ln}[Z_R(z,1/a,\mu;k,\Lambda_{\rm QCD},m_0,d)]\nonumber \\
& +\begin{cases}
\text{ln}\big[\tilde{h}^{\rm pert}_{\overline{{\text{MS}}}}(z,\mu)\big]  & \text{if } z_0 \le z \le z_1\\
g(z)-m_0z & \text{if } z_1<z
\end{cases} ,
\label{eq:self1_gluon}
\end{align}
where $k$, $\Lambda_{\text{QCD}}$, $m_0$, $d$ and $g(z)$ are treated as free parameters. We choose $z_0=0.15$ fm , $z_1=0.3$ fm and the maximum value of $z$ as 1 fm.  













\begin{figure}[h]
\includegraphics[width=.45\textwidth]{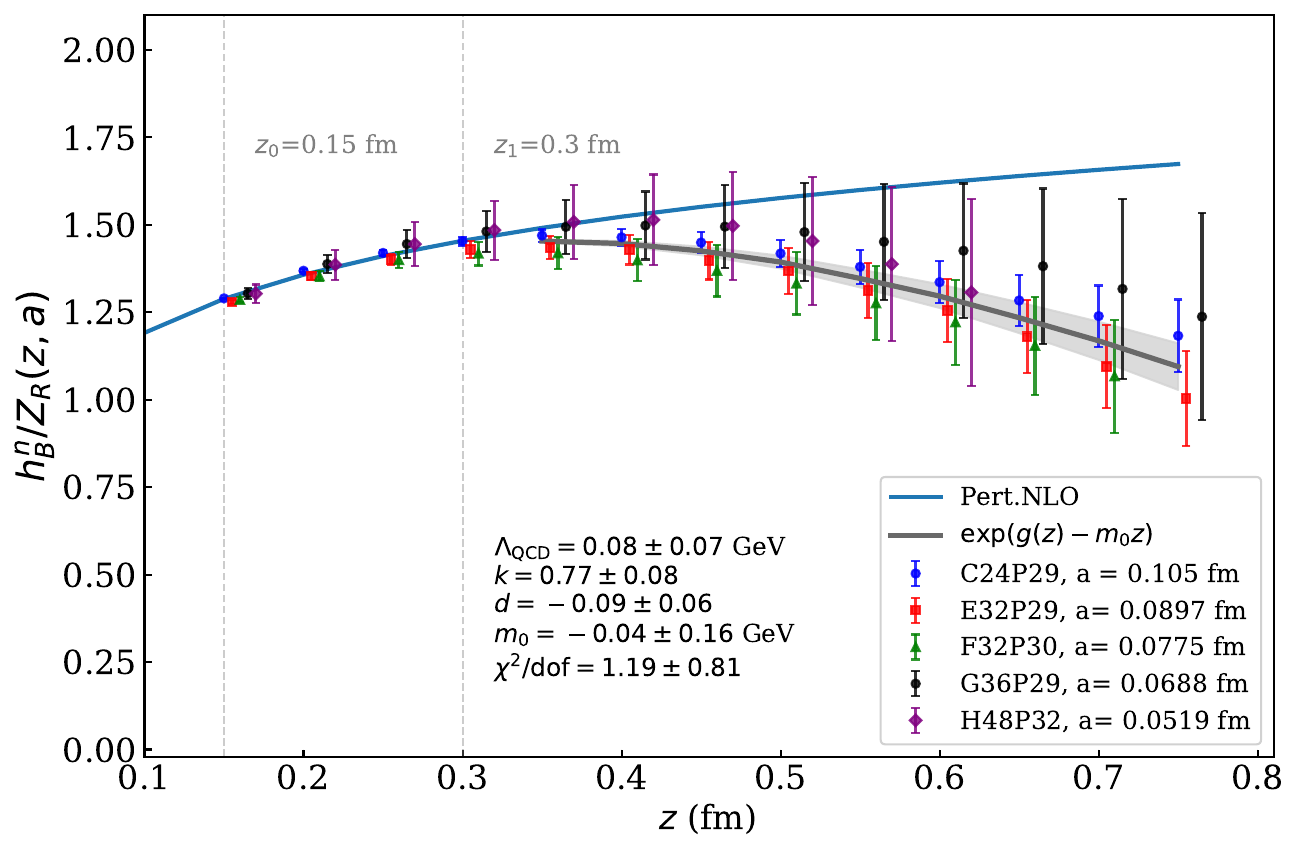}
\caption{ Comparison of renormalized matrix elements   with the perturbative one-loop $\overline{\text{MS}}$ result (black curve). Error bars represent the renormalized  matrix elements for each lattice spacing. $g(z)$ is the
residual from the fitting process and $\rm{exp}[g(z)-m_0 z]$ (gray band) indicates the $a$-independent renormalized matrix element. } 
\label{fig:fit_MSbar}
\end{figure}

Fig.~\ref{fig:fit_MSbar} shows a comparison of the
renormalized matrix elements with the perturbative one-loop $\overline{\text{MS}}$ results.
As shown in the figure, the renormalized matrix elements agree well with the continuum one-loop result at short distance while they deviate significantly at large distances due to non-perturbative infrared effects. 
The  fitting results of each parameter and the corresponding $\chi^2$/d.o.f. are also shown in the figure. 

\begin{figure}[!h]

 \centering
\includegraphics[width=.42\textwidth]{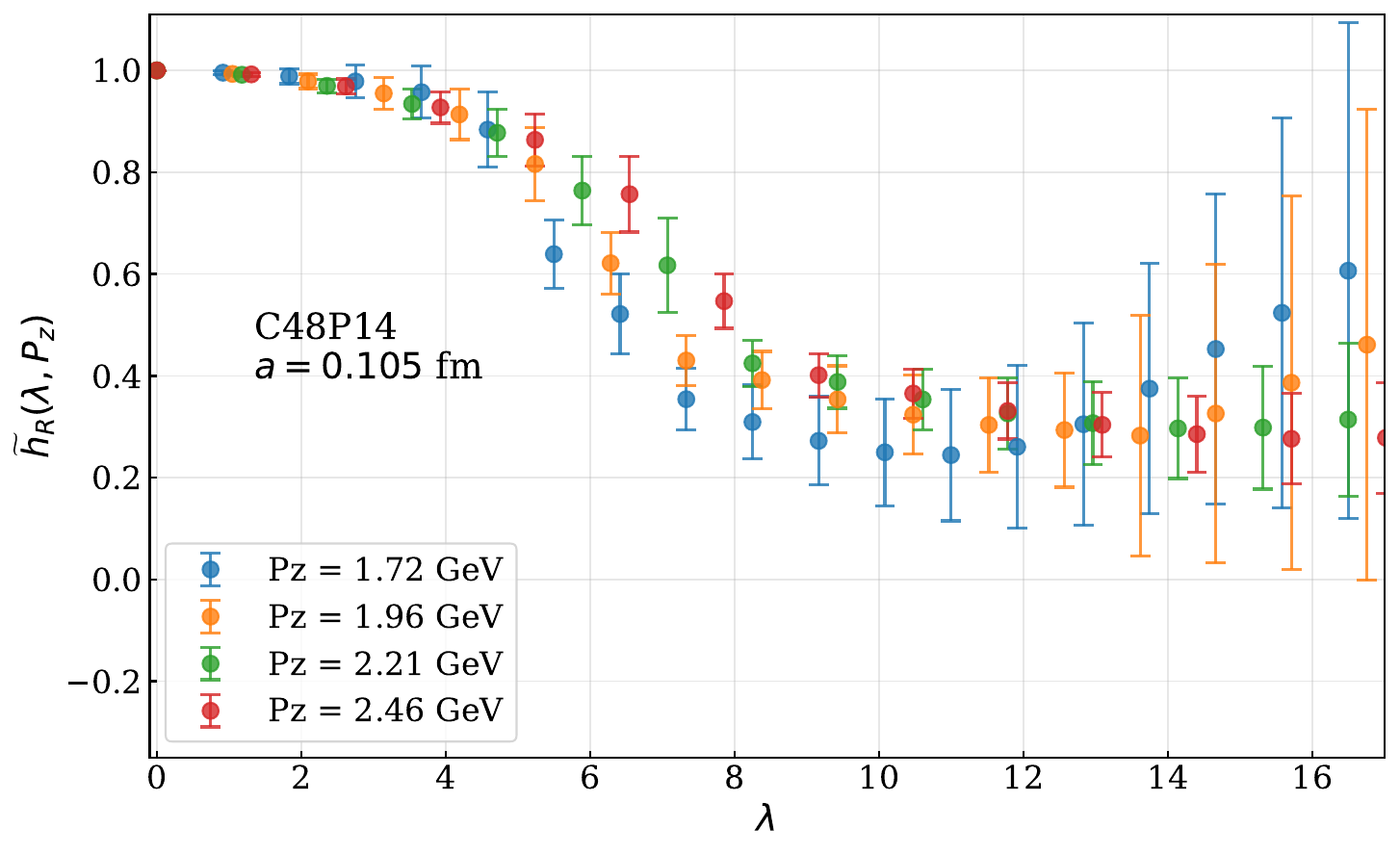}

\caption{ The renormalized matrix elements for the  C48P14 ensemble at the physical pion mass, as a function of $z$ at scale $\mu=2$~GeV, with different colors corresponding to different nucleon momenta.}
\label{fig:ren_ma_C48P14}
\end{figure}


The renormalized matrix elements for the ensemble C48P14 are shown in Fig.~\ref{fig:ren_ma_C48P14}, with different colors denoting different nucleon momenta. To suppress power corrections of order $1/P_z^2$, only results with $P_z>1.7~\mathrm{GeV}$ are preserved. 

\begin{figure}[!h]

 \centering
\includegraphics[width=.42\textwidth]{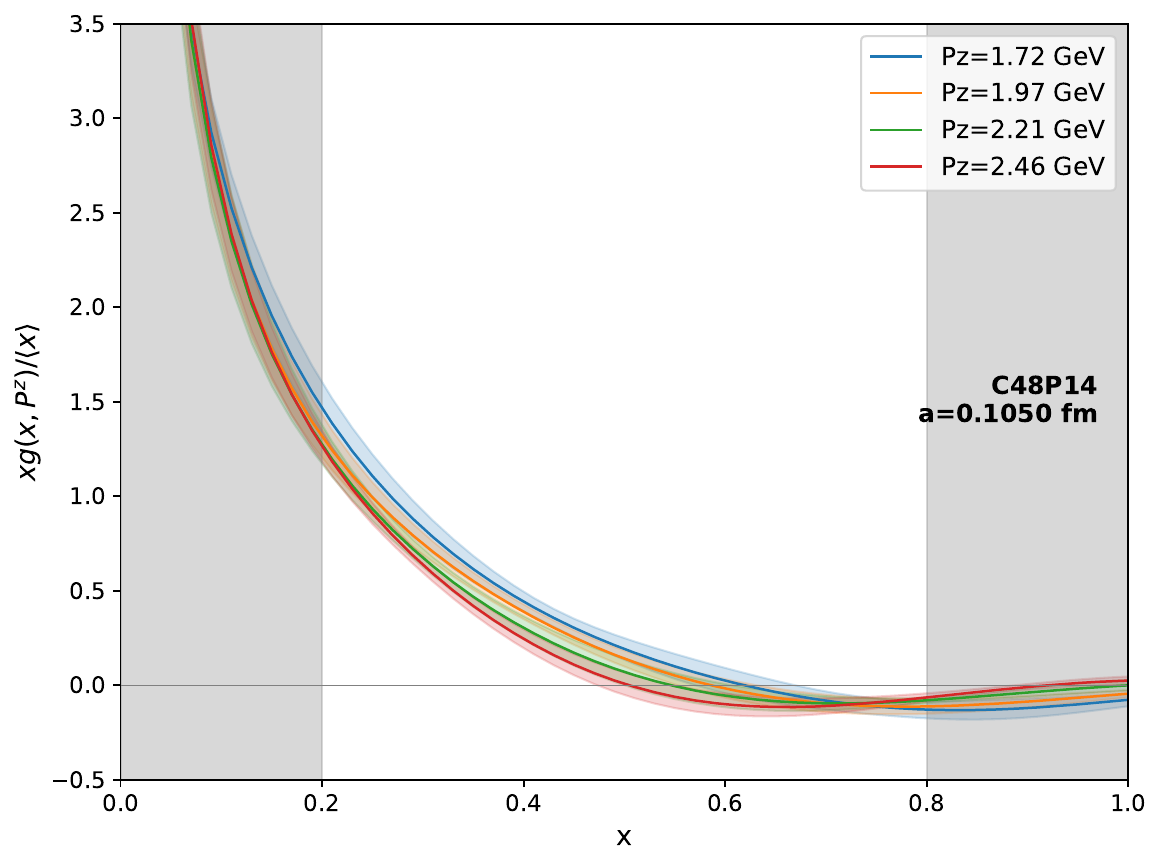}
\caption{Results of the light-cone PDF $xg(x)/\langle x \rangle$ for the C48P14 ensemble. The colored bands illustrate the results obtained at different momenta.  The gray bands denote the regions where LaMET results are expected to be unreliable.
}  
\label{fig:lc_mom_a_dependent_C48P14}
\end{figure}


The uncertainties of renormalized matrix elements become worse with increasing $\lambda= z P_z$. However, a Fourier transform to momentum space requires the quasi-light-front correlator at all distances. To resolve this issue, 
we follow Ref.~\cite{Ji:2020brr} and adopt an
extrapolation with the following form in the large $\lambda$ region:

\begin{align}
	  \tilde{h}_R(\lambda) &= l_1 \lambda ^{-a_1} e^{-\lambda/\lambda_0},
	  \label{eq:lambda_extrap}
\end{align} 
where $\lambda ^{-a_1}$ is associated with a power law behavior of the unpolarized PDFs in the end point region, and the exponential factor accounts for the expectation that the correlation length $\lambda_0$ of the correlation function is finite at finite momentum~\cite{Ji:2020brr}.  Examples of large $\lambda$ extrapolation can be found in the supplemental material ~\cite{supplemental}.

{\em Light-cone PDF:}
After the $\lambda$ extrapolation, we can perform a Fourier transform and obtain the $x$-dependent quasi-PDF in momentum space, and then 
extract the light-cone PDF  through perturbative matching. The detailed matching formula can be found in the supplemental material~\cite{supplemental}.


The light-cone PDF obtained after perturbative matching still contains power $P_z$ dependence, lattice spacing dependence and pion mass dependence, which has to be removed by  a combined extrapolation. 

\begin{align}
xg(x,P_z,a)&=xg_0(x)+a^2f(x) \\ \nonumber &+\frac{d(x)}{P_z^2}+k(x)(m_{\pi}^2-m_{\pi,phy}^2),
    \label{eq:extrap_form1}
\end{align}
where $xg(x,P_z,a)$ represents the PDF obtained on different ensembles and at different momenta, and  $xg_0(x)$ represents the light-cone PDF extrapolated to  infinite momentum limit, continuum limit and physical pion mass. The
${d(x)}/{P_z^2}$ term  accounts for the leading higher-twist contribution and $k(x)(m_{\pi}^2-m_{\pi,phy}^2)$ denotes the pion mass dependence. The physical pion mass $m_{\pi,phy}$ is taken to be $135~\mathrm{MeV}$. The \(a^2 f(x)\) term captures the discretization errors: theoretically the gluon field in the operator is free of \(O(a)\) effects since the gluon action is \(O(a^2)\) improved, and any \(O(a)\) contamination from the gauge link is purely imaginary and cannot contribute to the real matrix element of the unpolarized gluon quasi-PDF. Possible higher-order corrections like the lattice spacing dependence of $d(x)$ are not statistical significant and are omitted to ensure a stable fit.  

To further examine the possible  discretization effects, we perform numerical tests by varying the extrapolation ansatz. Specifically, to investigate the possible contribution from an $O(a)$ term, we replace the $a^2f(x)$ term by an $af_1(x)$ term in the extrapolation formula. 
In addition, we examine the effect of the momentum-dependent discretization term $a^2P_z^2h(x)$, which represents a higher-order momentum-dependent discretization effect analogous to the $a^2P_z^4$ correction in the lattice dispersion relation. 
The extrapolation results obtained from these different ansatzes are combined using the Akaike information criterion (AIC)~\cite{Borsanyi:2020mff, Wang:2025nsd} to estimate the model-dependent systematic uncertainty and determine the final central value of the combined extrapolation. 
The detailed information of the AIC analysis and the combined extrapolation results are provided in the supplemental material~\cite{supplemental}.


Results for the light-cone PDF $xg(x)/\langle x \rangle$ for the ensemble C48P14 obtained by perturbative matching are presented in Fig.~\ref{fig:lc_mom_a_dependent_C48P14}, with $\langle x \rangle=\int_0^1 \text{d} x x g(x) $  representing the gluon momentum fraction.  
The gray bands mark $x<0.2$ and $x>0.8$ regions, where higher order power corrections $\Lambda^2_\mathrm{QCD}/(xP_z)^2$ and $\Lambda^2_\mathrm{QCD}/((1-x)P_z)^2$ may become significant and cannot be neglected. At $0.2<x<0.8$  regions the nucleon momenta employed in this work are sufficiently large to suppress these power corrections, ensuring the reliability of the LaMET expansion.
The colored bands illustrate the results obtained at different momenta. 

In the supplemental material~\cite{supplemental}, we show the light-cone PDFs for other ensembles and detailed information of combined extrapolation. 

Systematic uncertainties associated with the large-$\lambda$ extrapolation, perturbative matching, combined extrapolation, and the choice of $z_s$ are included in this work.

The uncertainty from the large-$\lambda$ extrapolation is estimated by varying the fitting ranges and fitting forms and taking the differences from the default results as systematic uncertainties. The uncertainty associated with higher-order corrections in the perturbative matching is estimated by varying the renormalization scale from $2$ to $4$ GeV. The uncertainty due to the choice of $z_s$ in the hybrid renormalization scheme is estimated by taking the difference between the results obtained with $z_s=0.3$ fm and $z_s=0.2$ fm.
The uncertainty from the combined extrapolation is estimated by doing AIC analysis for different extrapolation forms. The detailed analysis of systematic uncertainties can be found in  the supplemental material~\cite{supplemental}.

\begin{figure}[h]
\includegraphics[width=.45\textwidth]{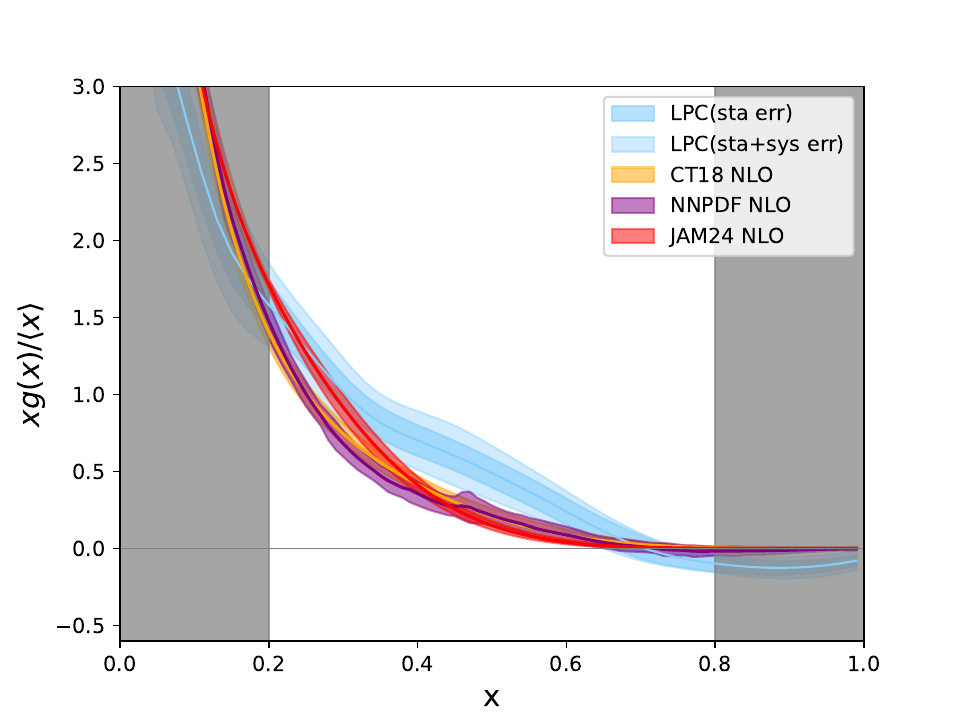} \\
\caption{LPC prediction of unpolarized gluon PDF in the continuum limit, infinite momentum limit and physical point $xg(x)/\langle x \rangle$ in comparison with  global fit results. The deep blue band
indicates statistical uncertainties of our result, while the light blue band represents
the combined effects of both statistical and systematic
uncertainties. The gray bands denote the regions where LaMET results are
expected to be unreliable.
}
\label{fig:mom_ext_sta}
\end{figure}

Our final results for the unpolarized PDF in the continuum, infinite momentum limit and physical pion mass are plotted  in  Fig.~\ref{fig:mom_ext_sta}. The deep blue band
indicates statistical uncertainties of our result, while the light blue band represents
the combined effects of both statistical and systematic uncertainties. 
We compare our results with CT18 NLO~\cite{Hou:2019qau}, NNPDF NLO~\cite{NNPDF:2017mvq} and JAM24 NLO~\cite{Anderson:2024evk} global fit , 
 and find them to be consistent with each other within 2 $\sigma$.
We observe that the theoretical and experimental results in the large-$x$ region are both  close to zero, indicating the suppression of the gluon PDF at large $x$. 
The large uncertainty of our final results is primarily due to the simultaneous extrapolation to the continuum, infinite momentum physical pion mass.  This process, which involves the \( 1/P_z^2 \) term, demonstrates the non-trivial systematic effects that must be conservatively treated.


{\em Summary:}
We have presented a lattice QCD calculation of the unpolarized gluon PDF of the nucleon using LaMET. Our calculation was
performed on ensembles with five lattice spacings ranging from $0.105$ fm to $0.0519$ fm and  pion mass ranging from $136$ MeV to $317$ MeV. With high statistics, we have used multiple source-sink separations to control the excited-state contamination and applied the state-of-the-art renormalization, matching, and extrapolation. The distillation method and momentum smearing technique were employed to enhance signal over noise ratio. We also extrapolated the result to the continuum limit , infinite momentum limit and physical pion mass. Our final result is consistent with the results of global fits to experimental data within 2 $\sigma$. 

Future improvements can be pursued in several directions. 
The renormalization and matching procedures could be refined by implementing renormalization-group resummation~\cite{Zhang:2023bxs} and leading-renormalon resummation~\cite{Su:2022fiu}. Developing these methods may extend the effective range of LaMET and reduce the scale dependence of our final results. 

{\it Note added:} While this paper was being finalized, Ref.~\cite{NieMiera:2025mwj} appeared which also computed the unpolarized gluon PDF of the nucleon using a different lattice setup based on gradient flow. In contrast to their calculation, we have performed an extrapolation to the infinite momentum limit, enabling a more direct connection to the unpolarized gluon PDF extracted from global fits. Nevertheless, the results are consistent within errors.  


\begin{acknowledgments}
\section*{Acknowledgments}
We thank the CLQCD collaboration for providing us with their gauge configurations with dynamical fermions~\cite{CLQCD:2023sdb,CLQCD:2024yyn}, which are generated on HPC cluster of ITP-CAS, IHEP-CAS and CSNS-CAS, the Southern Nuclear Science Computing Center(SNSC), the Siyuan-1 cluster supported by the Center for High-Performance Computing at Shanghai Jiao Tong University, and the Dongjiang Yuan Intelligent Computing Center. We thank Andreas Schaefer
for his constructive feedback on our manuscript. We are grateful to Yushan Su for the valuable discussion and comments. 
The numerical calculations in this paper were carried out at the Dongjiang Yuan Intelligent Computing Center,  the HPC cluster of ITP-CAS and  the Southern Nuclear Science Computing Center(SNSC).  This work is supported in part by National Natural Science Foundation of China (NSFC) under Grant No. 12293060, No. 12375080, No. 12293061, No. 12293062, No. 12175279, No. 12525504, No. 12293065, No. 12047503, No. 12435002, No. 12375080, No. 11975051 and 12447101, the Strategic Priority Research Program of the Chinese Academy of Sciences with Grant No. XDB34030301, No. XDB34030303 and No. YSBR-101, the Guangdong Major Project of Basic and Applied Basic Research No. 2020B0301030008,  the Education Integration Young Faculty Project of University of Chinese Academy of Sciences, the Ministry of Science and Technology of China under Grant No. 2024YFA1611004, CUHK-Shenzhen under grant No. UDF01002851, and a NSFC-DFG joint grant under Grant No. 12061131006 and SCHA 458/22 and the GHfund A No.\ 202107011598. 
FY is supported by the
U.S. Department of Energy, Office of Science, Office of
Nuclear Physics through Contract No. DE-SC0012704,
and within the framework of Scientific Discovery through
Advanced Computing (SciDAC) award Fundamental Nu-
clear Physics at the Exascale and Beyond.

\end{acknowledgments}

\bibliographystyle{apsrev}
\bibliography{ref}

\clearpage



\section*{Supplemental Material}\label{sec:supp}

\subsection{Theoretical framework}
The light-cone PDF of gluons (adjoint  representation) is defined as
\begin{align}
 xg(x,\mu)  & = \int \frac{dz^-}{2\pi P^+ } e^{-ix P^+ z^-} \nonumber \\  & \times \langle P|F^{a}_{+\mu}(z^-) 
U^{ab}(z^-,0) F^{b\mu}_{~~~~+}(0)(\mu)|P \rangle,
\end{align}\label{eq:lc-pdf_def}
where $z^\pm=\frac{1}{2}(z^0\pm z^3)$  is the space-time coordinate along the lightcone direction. The Wilson line along the light-cone direction 
\begin{align}
U(z^-,0)= \mathcal{P}\,{\rm exp}\big[ig\int_{0}^{z^-} d\eta^- A^+(\eta^-)\big]
\end{align}
 is introduced to ensure gauge invariance of the non-local bilinear operator , with  $A^{bc}_\mu=if^{abc}A_\mu^{a}$ representing  the gauge field.

According to LaMET, 
the light-cone PDF $yg(y, \mu)$ can be extracted from the quasi-PDF $x \tilde{g}(x, P_z)$ through a matching formula:
\begin{align}
   x \tilde{g}(x,P_z) = \int_{-\infty}^{\infty} \frac{dy}{|y|} C_{\text{hybrid}}\left(\frac{x}{y},\lambda_s, \frac{\mu}{y P_z}\right) y g(y, \mu) 
\nonumber\\+ \mathcal{O}\left(\frac{\Lambda_{\mathrm{QCD}}^2}{(x P_z)^2}, \frac{\Lambda_{\mathrm{QCD}}^2}{((1-x) P_z)^2}\right),
\label{lcmatching}
\end{align}
where $C_{\text{hybrid}}\left(\frac{x}{y},\lambda_s, \frac{\mu}{y P_z}\right)$ represents the perturbative matching kernel in momentum space. Refs.~\cite{Balitsky:2021qsr, Balitsky:2019krf} provide the coordinate-space perturbative matching  kernel up to next-to-leading order (NLO) in the ratio scheme. We subsequently derive  the momentum-space  counterpart through a direct Fourier transform, implementing the methodology detailed in Ref.~\cite{Yao:2022vtp}.
The corresponding perturbative kernel up to NLO in the hybrid and ratio scheme reads~
\begin{widetext}
\begin{align}\label{eq:mommatchingkernelratio}
 C_{\text{ratio}}&\left(\xi,\frac{\mu}{yP_z}\right)=\delta\left(1-\xi\right)
 +\frac{\alpha_s(\mu) C_A}{2\pi}
 \begin{cases}
\left[\frac{2(1-\xi+\xi^2)^2}{1-\xi}\ln\frac{\xi}{\xi-1}+\frac{11-28\xi+18\xi^2-12\xi^3}{6(1-\xi)}\right]_+ & \xi > 1 \\
\left[\frac{2(1-\xi+\xi^2)^2}{1-\xi}\left(-\ln\frac{\mu^2}{4y^2P_z^2}+\ln (\xi(1-\xi)) \right)-\frac{15-56\xi+102\xi^2-96\xi^3+48\xi^4}{6(1-\xi)}\right]_+ & 0<\xi<1 \\
\left[\frac{-2(1-\xi+\xi^2)^2}{1-\xi}\ln\frac{\xi}{\xi-1}-\frac{11-28\xi+18\xi^2-12\xi^3}{6(1-\xi)}\right]_+ & \xi<0 ,
\end{cases}
\end{align}

\begin{align}
C_{\text{hybrid}}(\xi,\lambda_s,\frac{\mu}{yP_z})=C_{\text{ratio}}&\left(\xi,\frac{\mu}{yP_z}\right)+\frac{\alpha_s(\mu) C_A}{2\pi}\frac{5}{6}\left
( -\frac{1}{|1-\xi|}+\frac{2\text{Si}((1-\xi)|y|\lambda_s)}{\pi(1-\xi)}\right)_+,
\end{align}
\end{widetext}
with $\xi=\frac{x}{y}$.

The quasi PDF, which can be directly calculated on the lattice, is defined as:

\begin{align}
x\tilde{g}(x,P_z)&=\int\frac{  P_z dz}{2\pi} e^{-ix zP_z } \tilde{h}_R(z,P_z) ,
\end{align}

where $\tilde{h}_R(z,P_z)=\langle P|O(z)|P \rangle_R$ is the renormalized matrix element  with   $O(z)$ representing a linear combination of gluon operator $M_{\mu\lambda,\nu\rho}(z)$, as shown in Eq.~\ref{ope_com}.


\subsection{Bare matrix element}
\subsubsection{Two-point and three-point correlators}
Bare matrix elements $\tilde{h}_B(z,P_z)$ can be obtained by calculating two-point and three-point correlators.
The nucleon's two-point correlator (2pt) is defined as 
\begin{align}
    C_\text{2pt}(P_z;t_{\text{sep}})=\sum_{t_0}\langle 0|N(P_z;t_0+t_{\text{sep}})N^{\dagger}(P_z;t_0)|0 \rangle 
\end{align}

with
\begin{align}
N(\vec{P};t)=\sum_{\vec{x}}e^{-i\vec{x} \cdot\vec{P}}P_+(u^TC\gamma_5 \gamma_t d)u(\vec{x};t)
\end{align}
representing  the nucleon operator~\cite{Zhang:2025hyo} .
The three-point correlator (3pt) is defined as
\begin{align}
    & C_\text{3pt}(P_z;z;t_\text{sep},t) \nonumber \\ &=\sum_{t_0} \langle 0|N(P_z;t_0+t_\text{sep})O(z; t_0+t)N^{\dagger}(P_z;t_0)|0 \rangle.
     \label{eq:3pt}
\end{align}

\begin{figure}[h]
\includegraphics[width=0.4\textwidth]{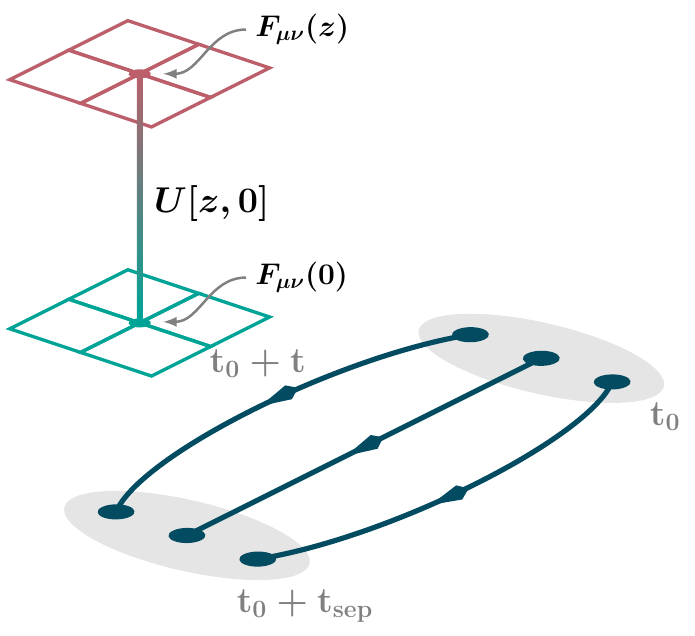}
\caption{ Illustration of the three-point correlator of gluon. }
\label{fig:gluon_3pt}
\end{figure}

On the lattice the three-point correlator can be obtained by calculating the disconnected insertion as shown in Fig~\ref{fig:gluon_3pt}.
It can be represented as:
 \begin{align}
     C_\text{3pt}\left(P_z;z;t_\text{sep},t\right) =\sum_{t_0}\left(O\left(z ; t_0+t\right)-\left\langle O\left(z; t_0+t\right)\right\rangle\right)\nonumber\\   \times\left(C_\text{2pt}\left(P_z; t_0+t_\text{sep}, t_0\right)-\left\langle C_\text{2pt}\left(P_z; t_0+t_\text{sep}, t_0\right)\right\rangle\right).
 \end{align}
 
\subsubsection{Discussion on the choices of HYP-smearing steps}

\begin{figure*}      
     \centering
     \includegraphics[width=0.3\linewidth]{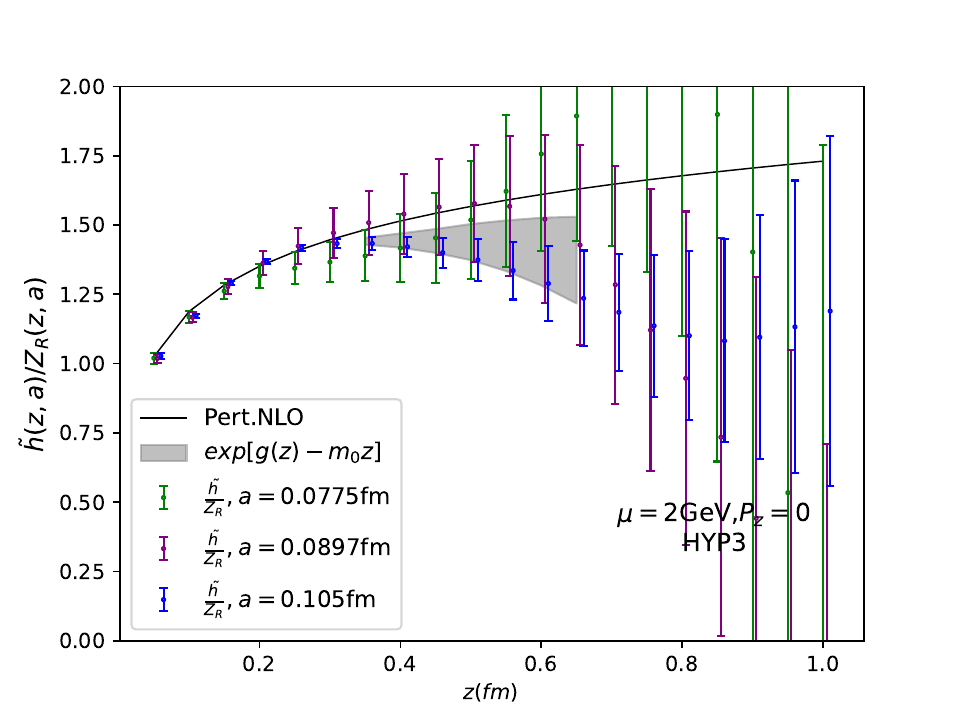}
      \includegraphics[width=0.3\linewidth]{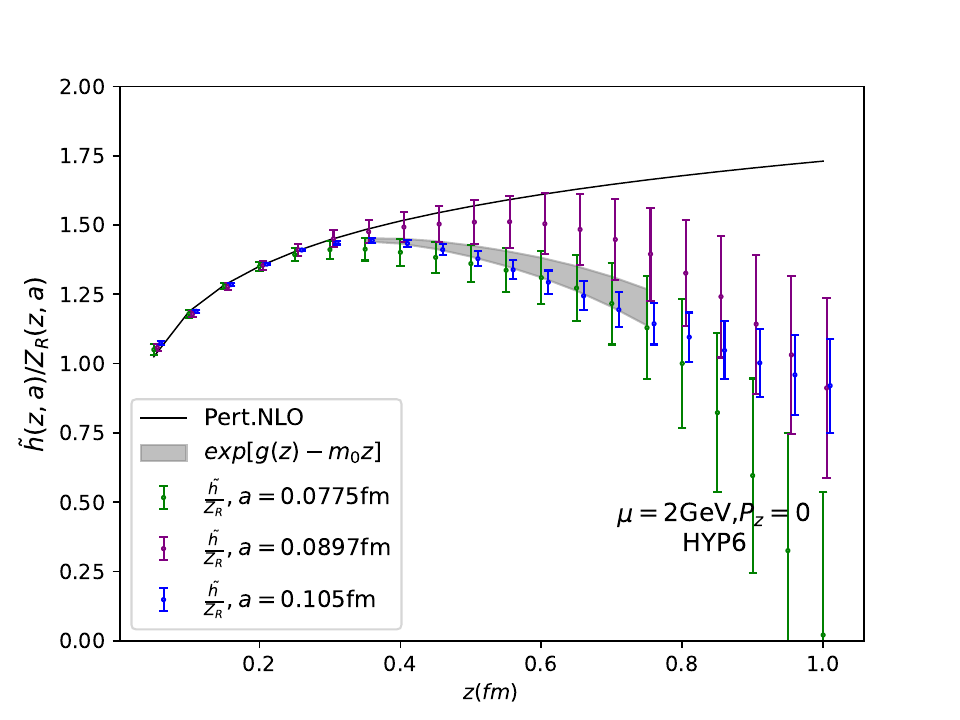}
       \includegraphics[width=0.3\linewidth]{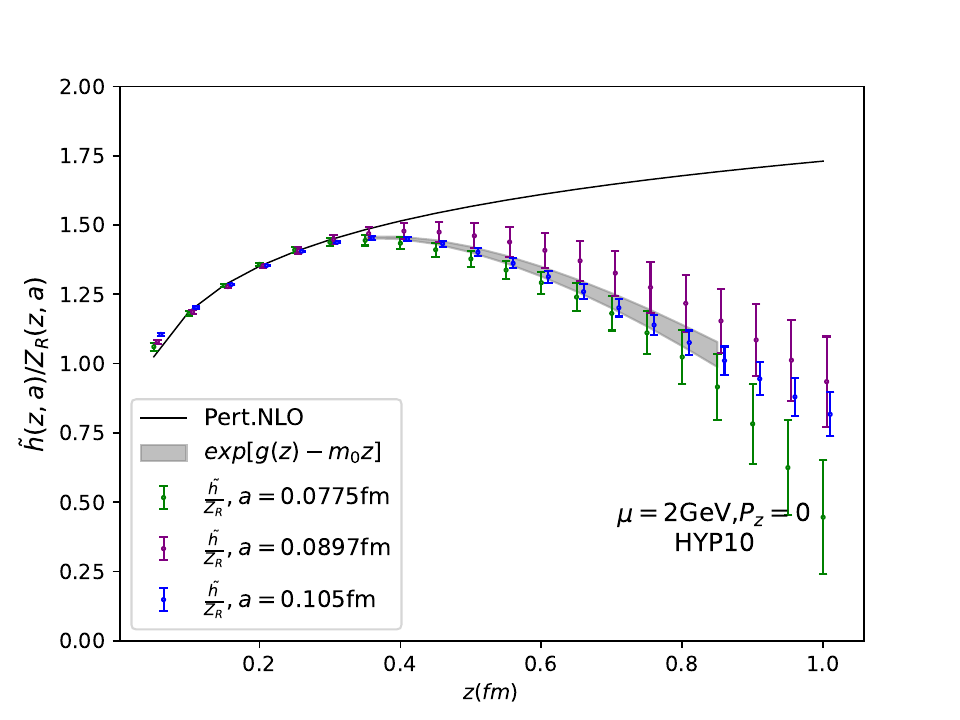}
  \caption{ Self-renormalized matrix elements under 3 times (left), 6 times (middle) and 10 times (right) HYP smearing.   The black curve represents the perturbative one-loop $\overline{\text{MS}}$ result . Error bars represent the renormalized  matrix elements for each lattice spacing. $g(z)$ is the
residual from the fitting process and $\rm{exp}[g(z)-m_0 z]$ (gray band) indicates the $a$-independent renormalized matrix element. } \label{fig:self_renME_HYP_times}
 \end{figure*}

In order to varify the increase of  HYP smearing times will increase the signal of renormalized matrix elements and in the meantime, does not distort the behaviour of renormalized matrix elements, we have carried out additional tests with different numbers of HYP-smearing steps. 
In Fig.~\ref{fig:self_renME_HYP_times}, we show the results of self-renormalized matrix elements at zero momentum obtained with 3, 6, and 10 steps of HYP smearing.  The corresponding self-renormalization factors are extracted from  the C24P29, E32P29, and F32P30 ensembles. As can be seen the signal of renormalized matrix elements are improved with increasing HYP-smearing times.

\begin{figure*}     
     \centering
     \includegraphics[width=0.4\linewidth]{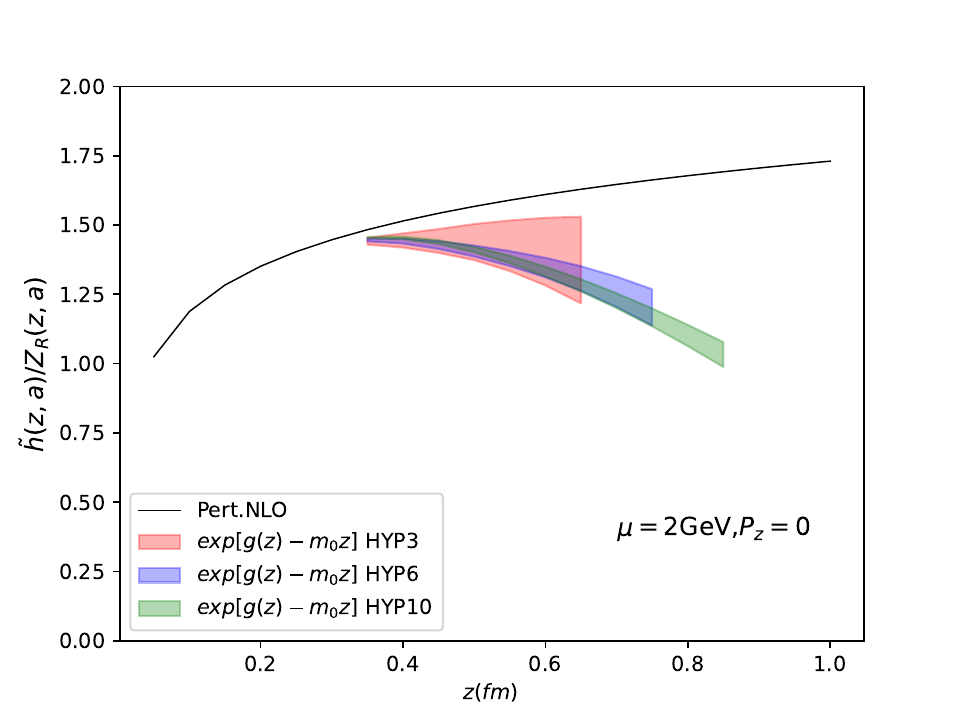}
          \includegraphics[width=0.4\linewidth]{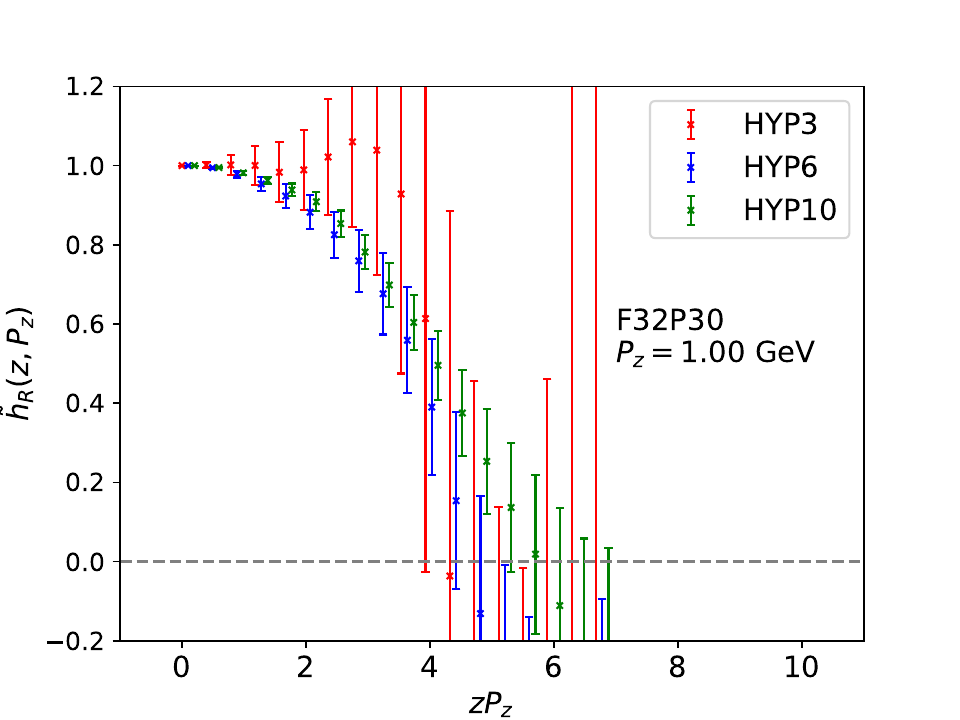}
     \caption{Self-renormalized matrix elements (exp$[g(z)-m_0z]$) at $P_z=0$ (left)  and hybrid-renormalized matrix elements of ensemble F32P30 at $P_z=1.00$GeV (right) under different HYP times.} \label{fig:renME_HYP_times_comp}
 \end{figure*}

For a more direct comparison, in Fig.~\ref{fig:renME_HYP_times_comp} we plot the renormalized matrix elements obtained with different numbers of HYP-smearing steps on the same figure.   In  the left panel, we compare the result of self-renormalized matrix elements through $\exp[g(z)-m_0 z]$ for different levels of HYP smearing.  In  the right panel  we compare the hybrid-renormalized matrix elements of ensemble F32P30 at $P_z=1.00$~GeV.  As shown in both plots, the  renormalized matrix elements of different HYP times are consistent with each others within uncertainties.

These tests indicate that, although HYP smearing improves the ultraviolet behavior of the bare matrix elements and the statistical quality of the signal,  the renormalized matrix elements do not show a significant dependence on the number of HYP-smearing steps used in our analysis. Therefore, within our current precision, we do not observe evidence that the level of HYP smearing employed in this work distorts the physical signal. 

\subsubsection{Fitting for bare matrix elements}
The effective matrix element can be extracted by first summing the ratio of the three-point and two-point correlator over the operator insertion time $t$  and then taking the difference with respect to the source-sink separation $t_{\mathrm{sep}}$:
\begin{align}
    \Delta R(P_z,z,t_{\mathrm{sep}})
    &=
    \frac{\sum_{t=t_r}^{t_{\mathrm{sep}}+1-t_r} C_{3\mathrm{pt}}(P_z,z;t,t_{\mathrm{sep}}+1)}
    {C_{2\mathrm{pt}}(P_z,t_{\mathrm{sep}}+1)}
    \nonumber\\
    &\quad -
    \frac{\sum_{t=t_r}^{t_{\mathrm{sep}}-t_r} C_{3\mathrm{pt}}(P_z,z;t,t_{\mathrm{sep}})}
    {C_{2\mathrm{pt}}(P_z,t_{\mathrm{sep}})},
\end{align}
here $n_r=t_r/a$ points near source and sink are excluded when summing over the operator insertion time $t$. In our analysis, this ratio is constructed from 3000 bootstrap samples of the three-point and two-point correlators. According to the Feynman-Hellmann theorem~\cite{Bouchard:2016heu}, the resulting effective matrix element can be decomposed into a ground-state contribution together with the leading excited-state contamination, with contributions from higher excited states neglected:
\begin{align}
    \Delta R(P_z,z,t_{\mathrm{sep}})
    \approx
    h_B(z,P_z)
    +
    c_1\, t_{\mathrm{sep}} e^{-\Delta E\, t_{\mathrm{sep}}}.
\end{align}
Here, \(h_B(z,P_z)\) denotes the bare matrix element. At large \(t_{\mathrm{sep}}\) , the excited-state contribution is exponentially suppressed and can be neglected. The bare matrix element can then be obtained from a constant fit,
\begin{align}
    \Delta R(P_z,z,t_{\mathrm{sep}})
    \approx
    h_B(z,P_z).
\end{align}

\begin{figure*}[hbp]
\includegraphics[width=1\textwidth]{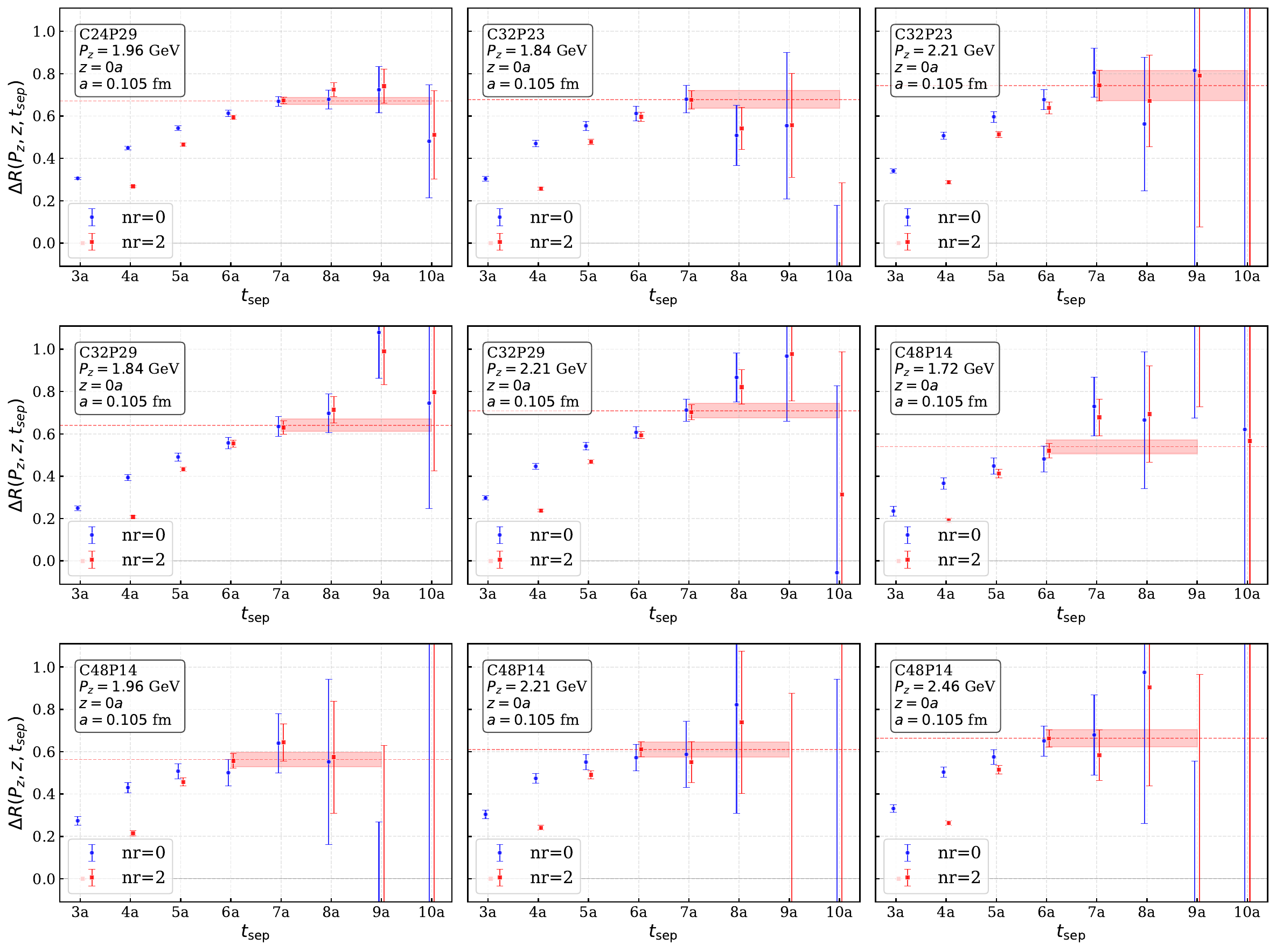}
\\
\caption{ The effective matrix element and the Feynman-Hellman fitting results for the  ensemble C24P29, C32P23, C32P29 and C48P14 with lattice spacings $a=0.105 $ fm . We  present the results at $z=0$ . The lattice data of the  effective matrix elements(error bars) and the fitting results of bare matrix elements(colored bands) are compared. and  $nr$ is the number of points removed from  both sides of each tsep.}
\label{fig:ratio_fit_FH_part1_a0}
\end{figure*}

\begin{figure*}[hbp]
\includegraphics[width=1\textwidth]{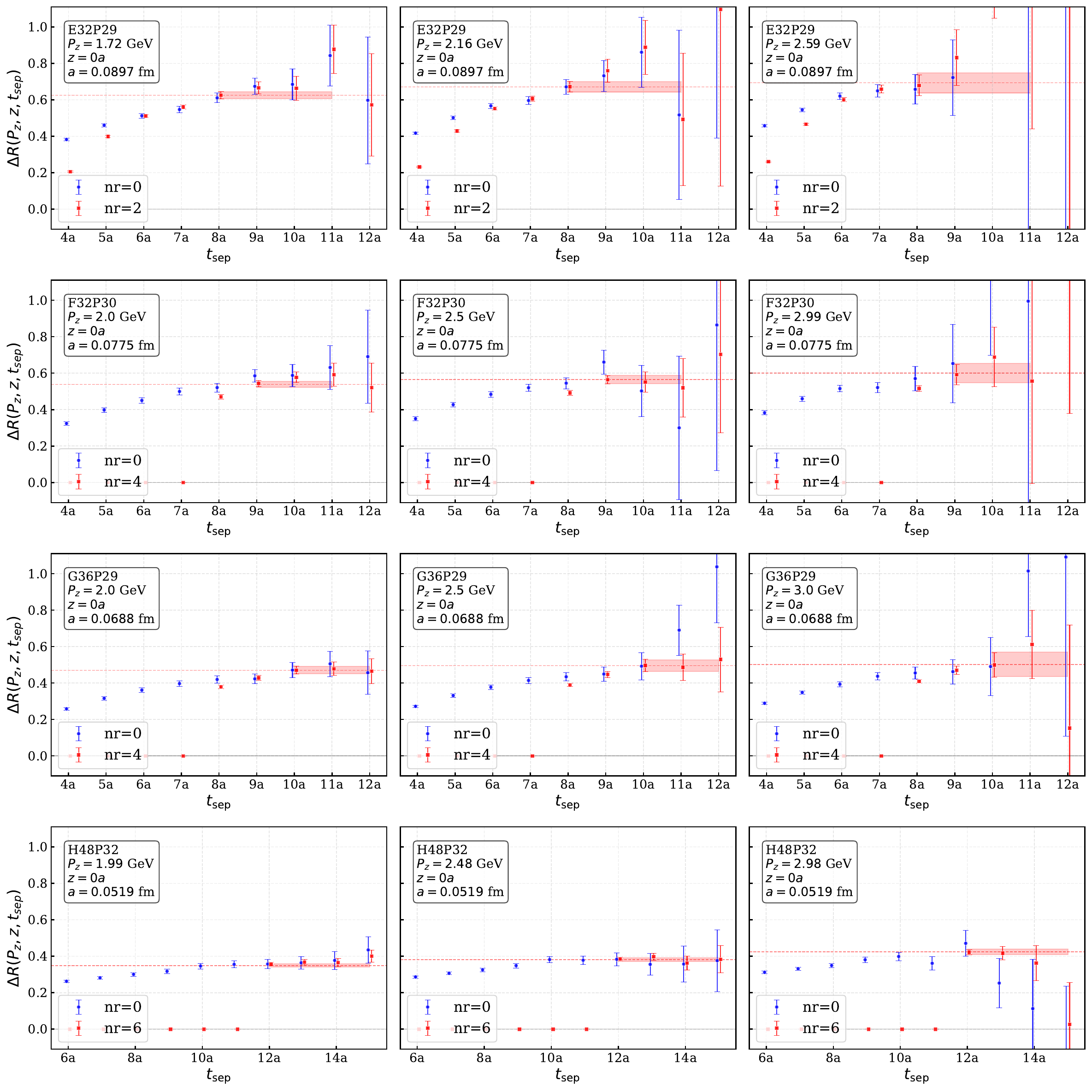}
\\
\caption{ The effective matrix element and the Feynman-Hellman fitting results for   ensemble E32P29, F32P30, G36P29 and  H48P32. We  present the results at $z=0$ . The lattice data of the  effective matrix elements(error bars) and the fitting results of bare matrix elements(colored bands) are compared, and  $nr$ is the number of points removed from  both sides of each tsep.}
\label{fig:ratio_fit_FH_part2_a0}
\end{figure*}

\begin{figure*}[hbp]
\includegraphics[width=1\textwidth]{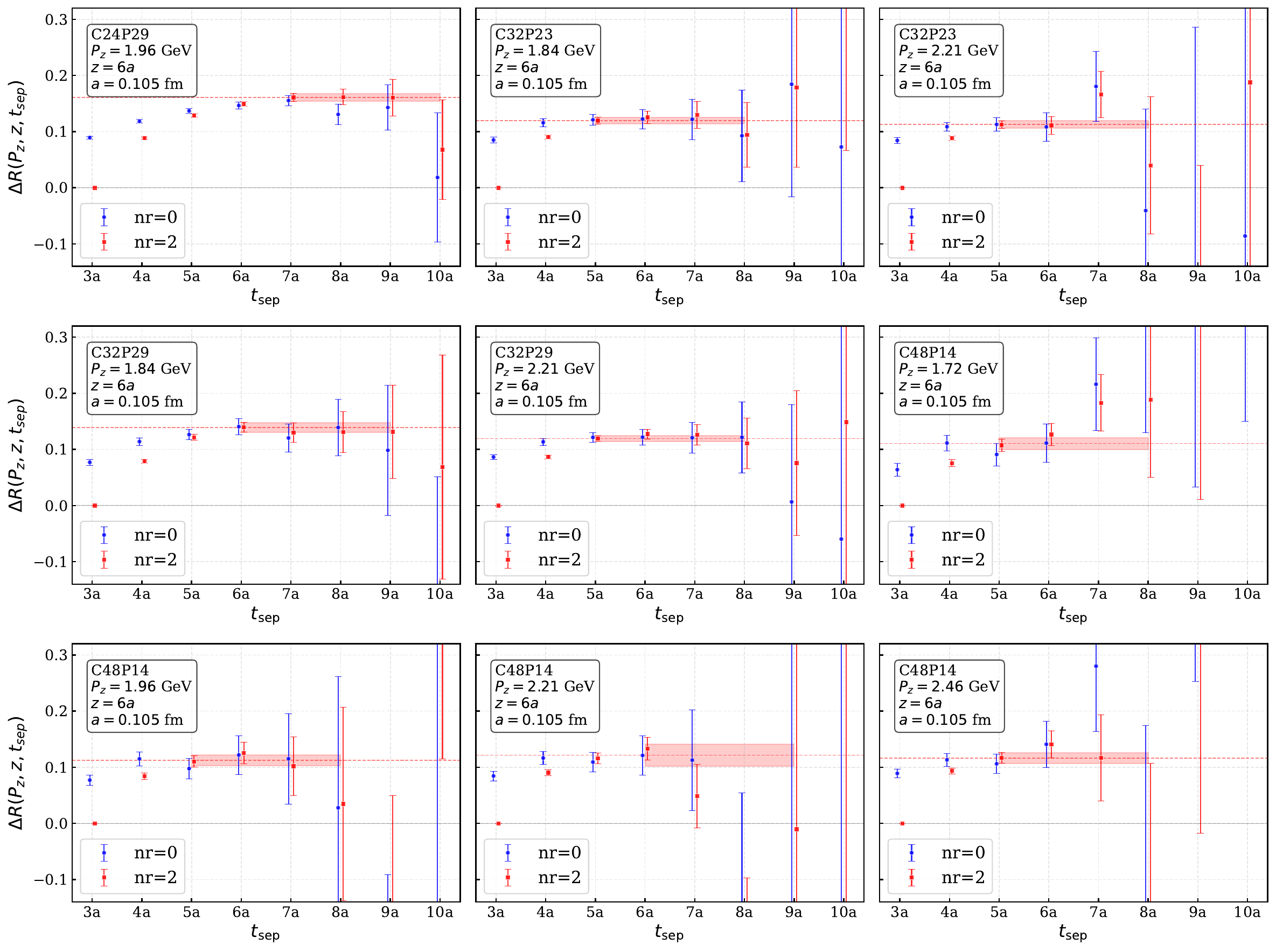}
\\
\caption{ The effective matrix element and the Feynman-Hellman fitting results for the  ensemble C24P29, C32P23, C32P29 and C48P14 with lattice spacings $a=0.105 $ fm . We  present the results at $z=6a$. The lattice data of the  effective matrix elements(error bars) and the fitting results of bare matrix elements(colored bands) are compared. and  $nr$ is the number of points removed from  both sides of each tsep.}
\label{fig:ratio_fit_FH_part1_a6}
\end{figure*}

\begin{figure*}[hbp]
\includegraphics[width=1\textwidth]{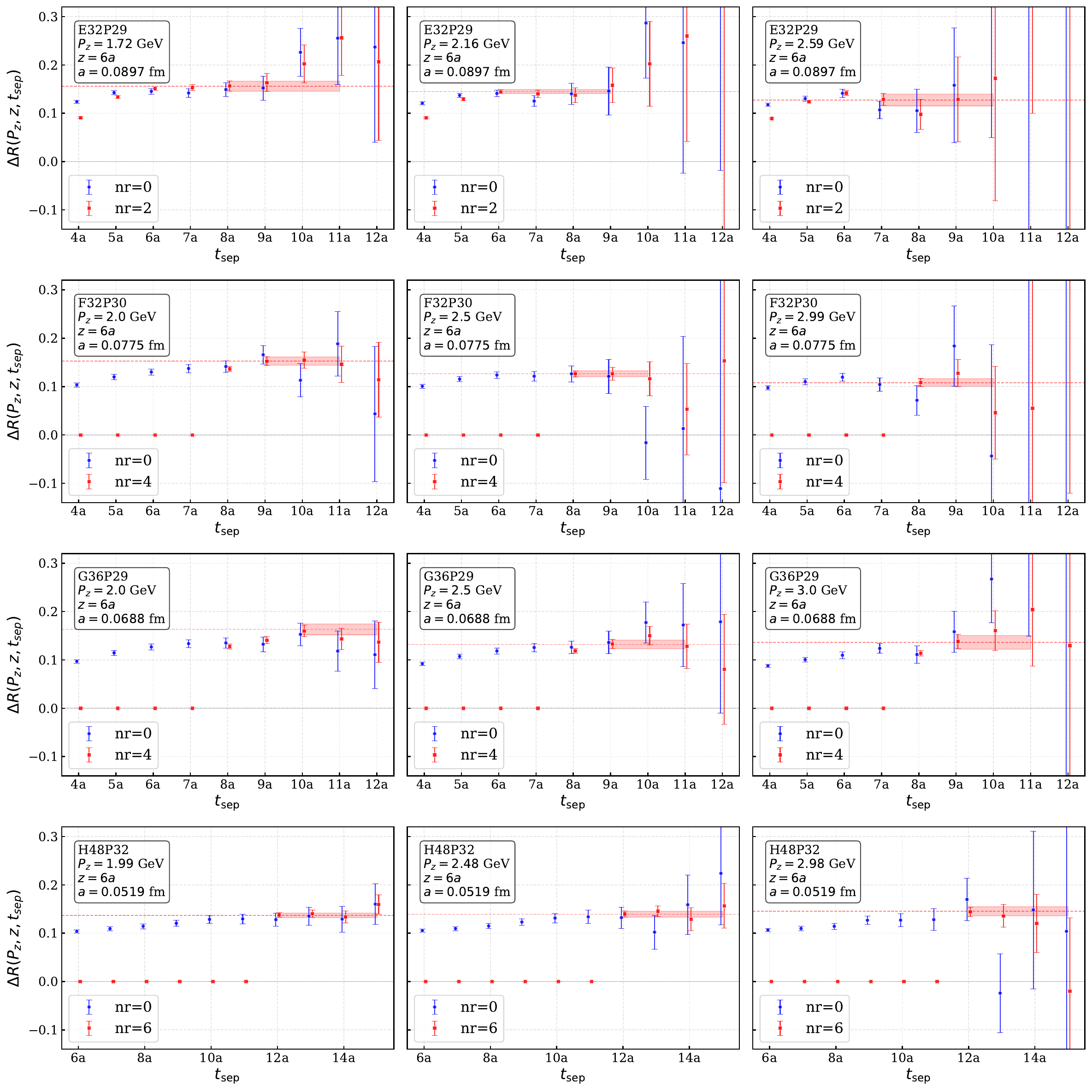}
\\
\caption{ The effective matrix element and the Feynman-Hellman fitting results for   ensemble E32P29, F32P30 , G36P29 and  H48P32. We present the results at $z=6a$. The lattice data of the  effective matrix elements(error bars) and the fitting results of bare matrix elements(colored bands) are compared, and  $nr$ is the number of points removed from  both sides of each tsep.}
\label{fig:ratio_fit_FH_part2_a6}
\end{figure*}

In  Fig.~\ref{fig:ratio_fit_FH_part1_a0}, Fig.~\ref{fig:ratio_fit_FH_part2_a0}, Fig.~\ref{fig:ratio_fit_FH_part1_a6},  and Fig.~\ref{fig:ratio_fit_FH_part2_a6}  we plot the  effective matrix element and the corresponding fitting results of bare matrix elements. We  present the
results at $z = 0$ and $z = 6a$. Error bars with different colors represent the result with different endpoint cut $n_r$ . We find that increasing $n_r$  makes the statistical errors of the extracted matrix elements smaller, but also increases the excited-state effects and makes the ground-state plateau appear later. Therefore, a suitable value of $n_r$ should be chosen to balance these effects. 
The red band indicates the fit obtained using our optimized choice of $n_r$. As can be seen, the corresponding fit range ensures the appearance of a stable plateau while significantly reducing the statistical uncertainty compared with the $n_r=0$ result, represented by the blue band. 
The momentum smearing parameters, optimized end-point cut value $n_r$   are shown in table~\ref{Tab:qgME}.  We individually optimize the fitting interval for each ensemble and each spatial shift 
$z$ to guarantee that the corresponding  $\chi^2/\rm{d.o.f}$ of all fits are approximately around unity. By adjusting the optimized fitting range,  we aim to obtain the result with the smallest possible error while  maintaining the reasonable $\chi^2/\rm{d.o.f}$ .  This approach allows us to postpone the fitting range for the 
$\lambda$ extrapolation and thereby reduce the model dependence of our final results.   



\begin{figure*}
\includegraphics[width=1\textwidth]{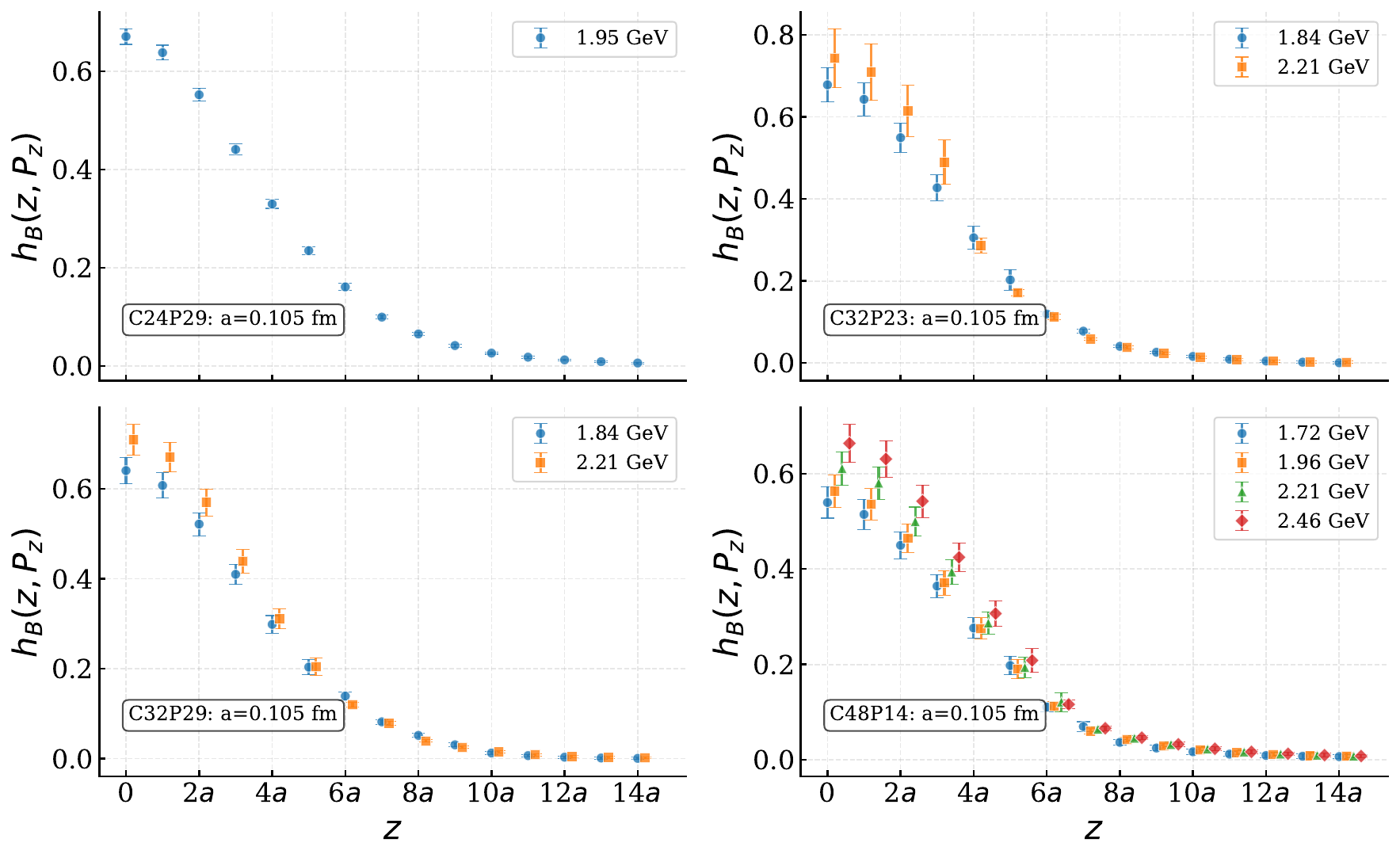}
\caption{ Bare matrix elements   obtained by fitting the ratio of three-to-two-point  for the  ensembles (C24P29, C32P23, C32P29, C48P14) with lattice spacings $a=0.105 $ fm . The results for different momenta are represented by different colors.}
\label{fig:hB_2}
\end{figure*}

\begin{figure*}
\includegraphics[width=1\textwidth]{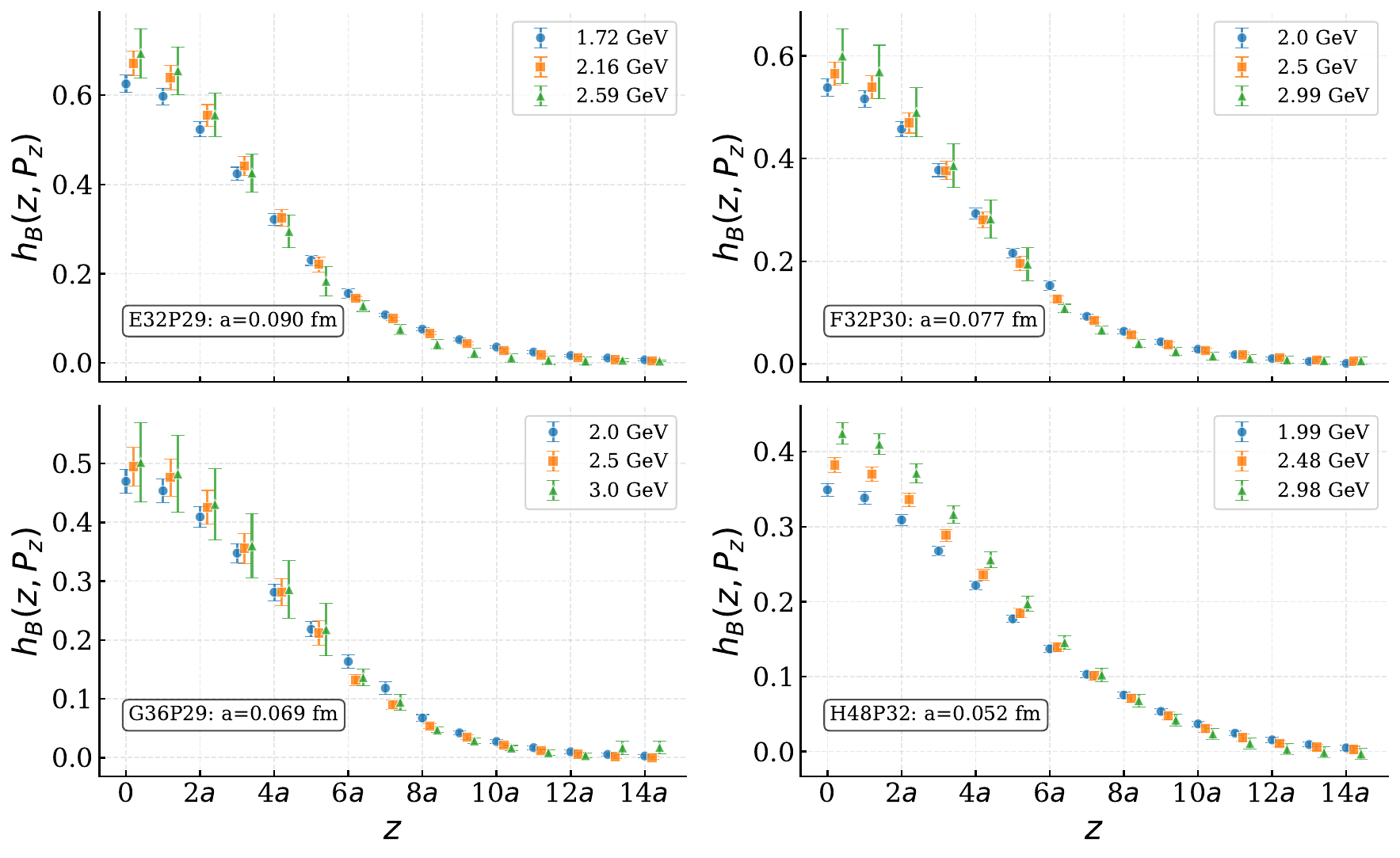}
\caption{Bare matrix elements   obtained by fitting the ratio of three-to-two-point  for the  ensembles (E32P29, F32P30, G36P29, H48P32) corresponding lattice spacings. The results for different momenta are represented by different colors.}
\label{fig:hB_3}
\end{figure*}

\

\renewcommand{\arraystretch}{1.5}
\begin{table}
\footnotesize
\centering
\begin{tabular}{ccccc}
\hline
\hline
Ensemble ~&$P_z$(GeV) ~&$n_{ \rm sm}$    ~& $\mathrm{n_{ex}}$ \\
\arrayrulecolor{black}

\hline

C24P29  ~& 0 ~& 0 ~&  $0$           \\
  ~     &1.95 ~& 2 ~&  2     \\
\hline
C32P29  ~ &1.84, 2.21 ~& 2 ~&  2     \\
\hline
C32P23     &1.84, 2.21 ~& 3 ~&    2   \\

\hline
C48P14    &1.72, 1.96, 2.21, 2.46 ~& 4 ~&      2 \\
\hline
E32P29  ~& 0 ~& 0 ~&  $0$           \\
  ~     &1.72, 2.16, 2.59 ~& 2 ~&       2 \\
\hline
F32P30  ~& 0 ~& 0 ~&  $0$           \\
  ~     &2.0, 2.5, 2.99 ~& 2 ~&     4  \\
\hline
G36P29  ~& 0 ~& 0 ~&  $0$           \\
  ~     &2.0, 2.5, 3.0 ~& 2 ~&     4  \\
\hline
H48P32  ~& 0 ~& 0 ~&  $0$           \\
  ~     &1.99, 2.48, 2.98 ~& 2 ~&    6   \\

\arrayrulecolor{black}
\hline

\end{tabular}
 \caption{The  nucleon's momentum $P_z$, the momentum smearing parameter $n_{\rm sm}$ and $n_\text{ex}$, the number of points excluded for both sides of  each  $t_{\mathrm{sep}}$.}
 \label{Tab:qgME}
\end{table}

\subsection{Renormalized matrix elements across all ensembles}
\begin{figure*}[t]
\centering
\begin{minipage}[t]{0.48\textwidth}
\centering
\includegraphics[width=\linewidth]{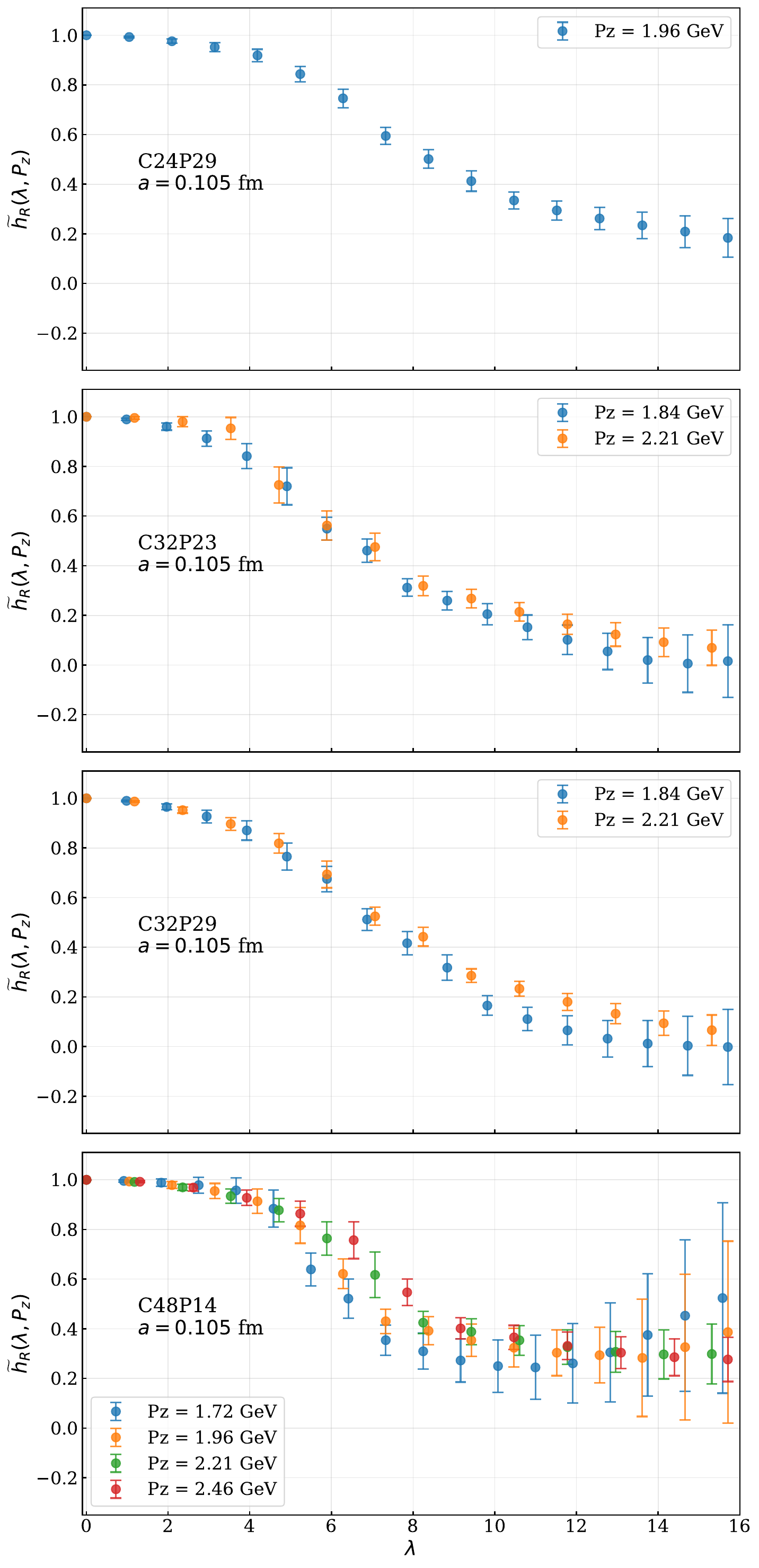}
\end{minipage}
\hfill
\begin{minipage}[t]{0.48\textwidth}
\centering
\includegraphics[width=\linewidth]{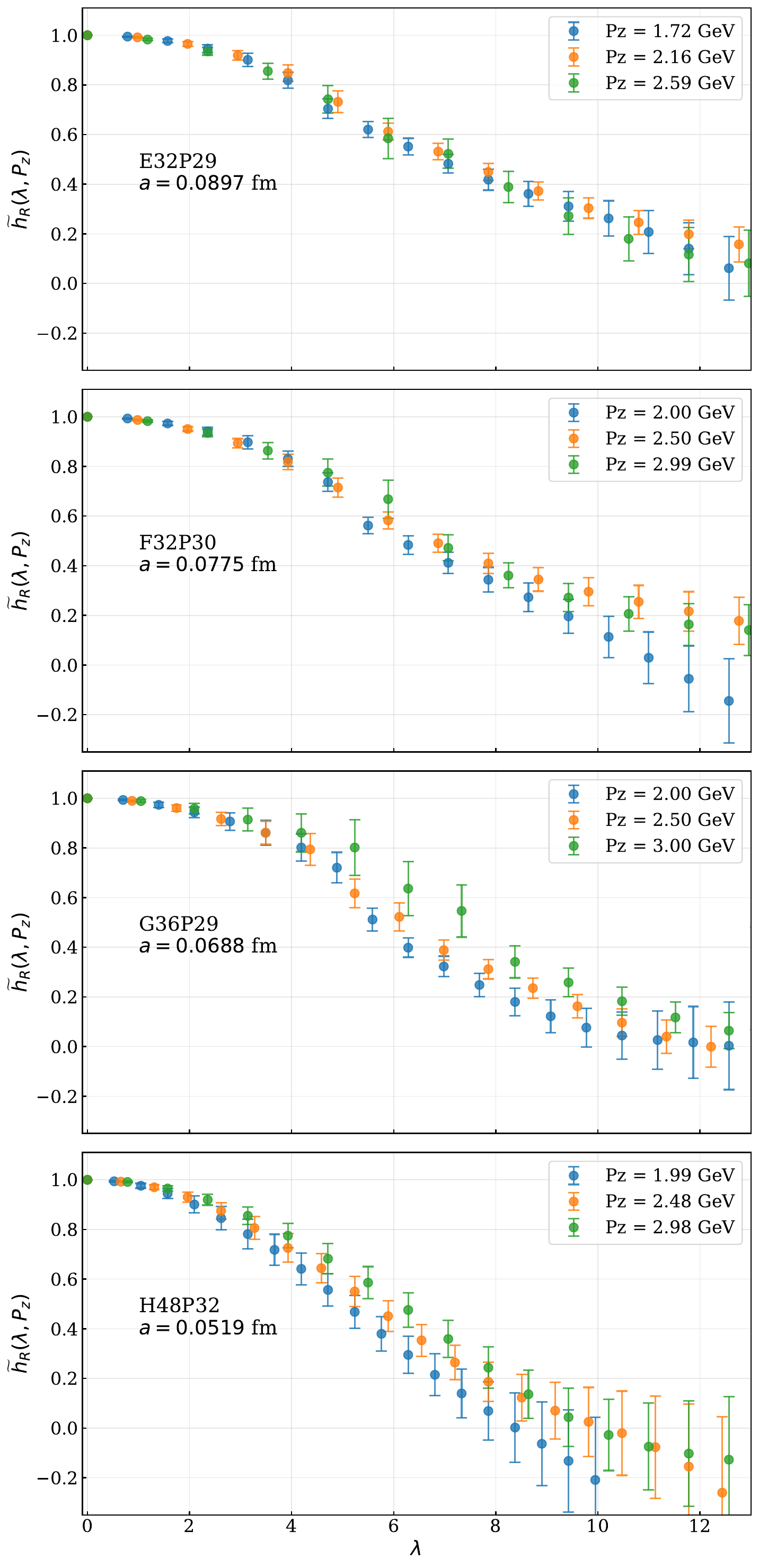}
\end{minipage}
\caption{
 The renormalized matrix elements for all ensembles as a function of $\lambda=zP_z$ at scale $\mu=2$~GeV, with different colors corresponding to different nucleon momenta.
}
\label{fig:ren_ma}
\end{figure*}

The renormalized matrix elements for all ensembles  are shown in Fig.~\ref{fig:ren_ma}, with different colors denoting different nucleon momenta. To suppress power corrections of order $1/P_z^2$, only results with $P_z>1.7~\mathrm{GeV}$ are preserved. As can be seen from the figures, all ensembles exhibit a similar convergence pattern as the nucleon momentum increases.





\subsection{Large-$\lambda$ extrapolation}
\begin{figure*}
\includegraphics[width=1\textwidth]{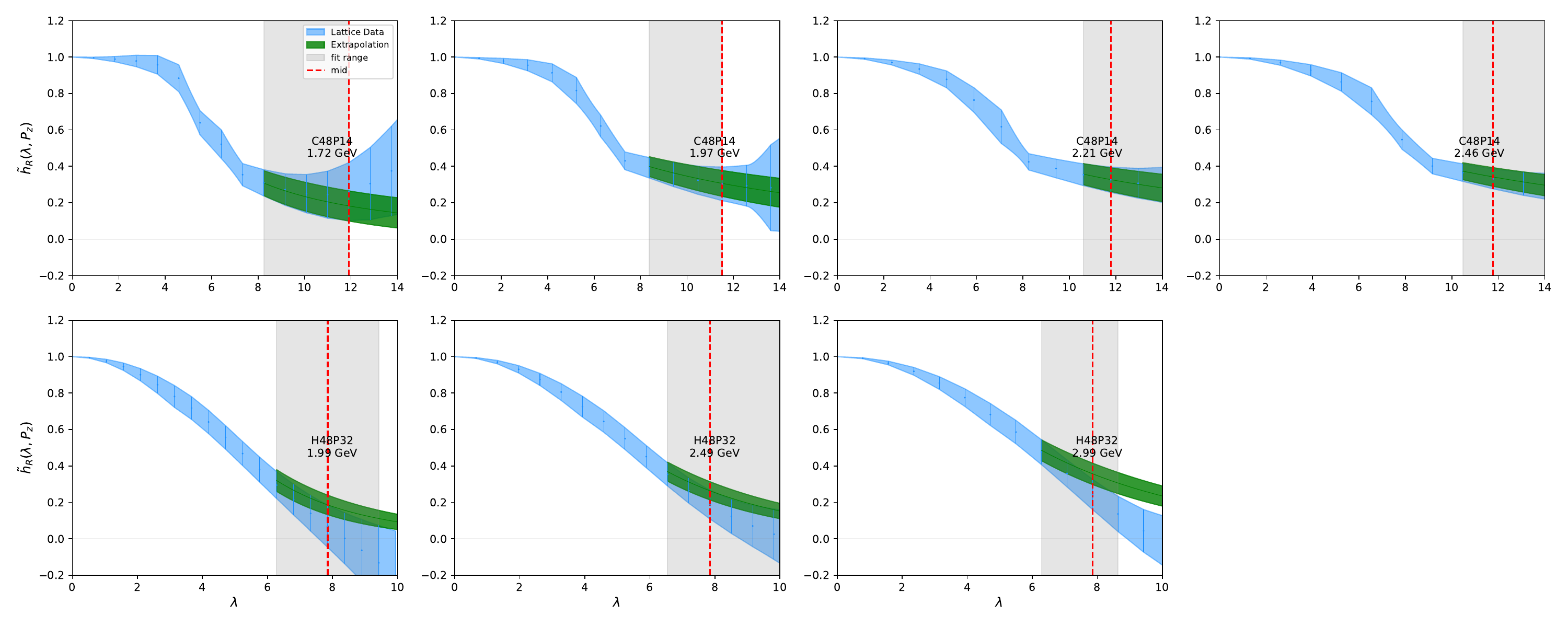}
\caption{ The large-$\lambda$ extrapolation of renormalized matrix elements, taking  ensemble C48P14 and  H48P32 as  examples. We
 present the results at $\mu = 2$ GeV, The blue band and green band correspond to lattice data and extrapolated results, respectively. The fitting range for Eq.~\ref{eq:lambda_extrap} is highlighted by the gray shaded region, and the red dashed line  in the plot marks the junction ($\lambda \geq 8$) between lattice data and extrapolated results during Fourier transformation.}
\label{fig:lambda_extrap_part1}
\end{figure*}

Performing the Fourier transform to momentum space requires the renormalized matrix elements across the entire coordinate space. A simple truncation would introduce oscillation into the momentum-space results. This issue can be resolved by performing extrapolation using Eq.~\ref{eq:lambda_extrap} to complete the results in the large-$\lambda$ region. In  Fig.~\ref{fig:lambda_extrap_part1}, we show the extrapolation of the renormalized matrix elements for different momenta, taking  ensemble C48P14 and  H48P32 as  examples. The fitting range for Eq.~\ref{eq:lambda_extrap} is highlighted by the gray shaded region, and the red dashed line  in the plot marks the junction ($\lambda \geq 8$) between lattice data and extrapolated results during Fourier transformation.
As can be seen from the plots the extrapolation results agree well with lattice data in fit range. 

\subsection{Light-cone PDFs of all ensembles}
\begin{figure*}
\includegraphics[width=0.8\textwidth]{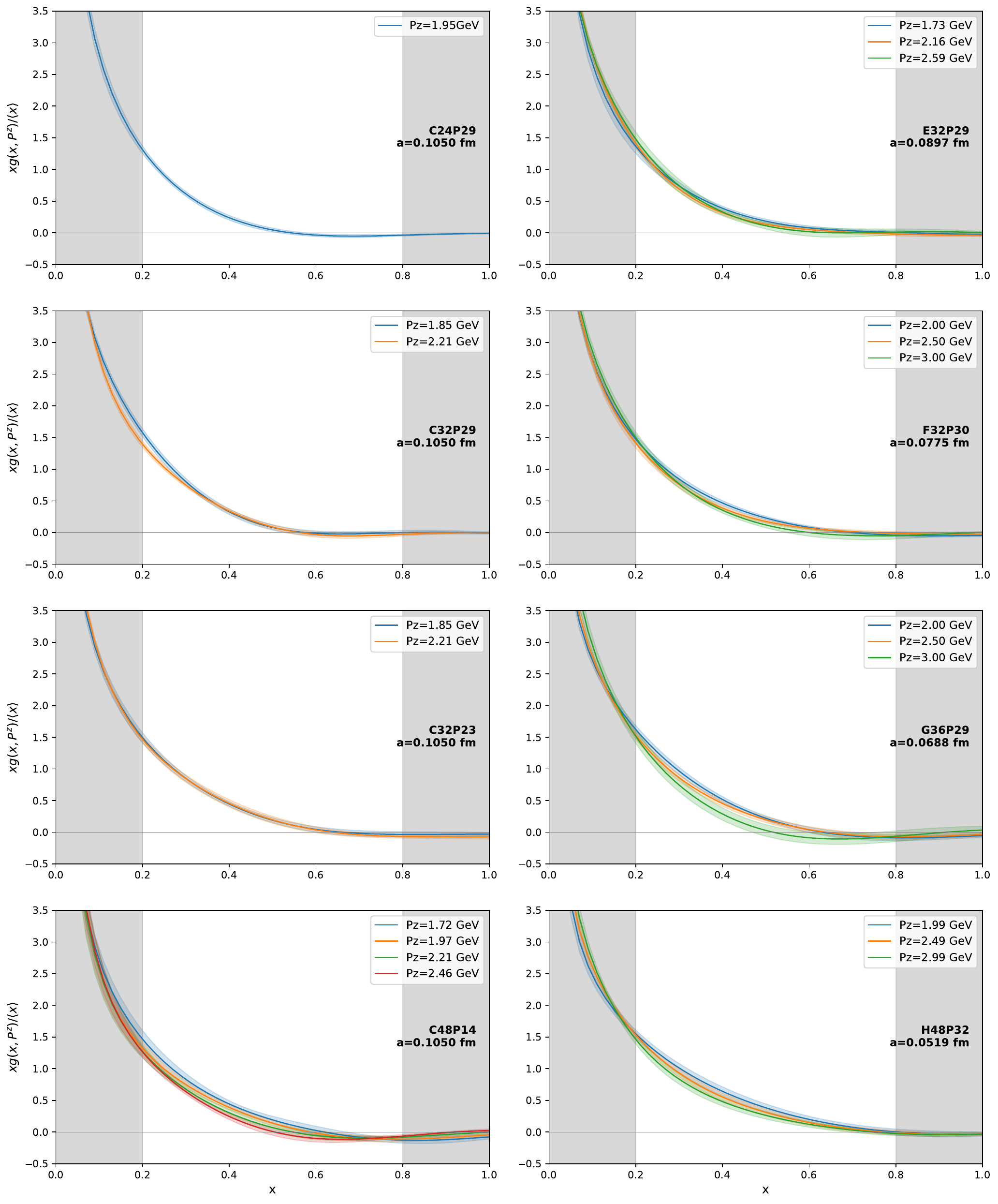}
\caption{ Results of the light-cone PDF $xg(x)/\langle x \rangle$ for all ensembles, with different colors  corresponding to different nucleon momenta. The gray bands denote the regions where LaMET results are expected to be unreliable.
 Only statistical uncertainties are shown.}
\label{fig:lc_mom_a_dependent}
\end{figure*}

Results for the light-cone PDF $xg(x)/\langle x \rangle$ for eight ensembles obtained by perturbative matching are presented in Fig.~\ref{fig:lc_mom_a_dependent}, with $\langle x \rangle=\int_0^1 \text{d} x x g(x) $  representing the gluon momentum fraction.  
The colored bands illustrate the results obtained at different lattice spacings and different momenta. As can be seen from the figures, the results exhibit similar trends as the nucleon momentum increases.


\subsection{Further discussion on combined extrapolation}
\label{subsection:discuss on combined extrap}

 As mentioned  in the main text we further examine the possible  discretization effects by repeating the extrapolation with several alternative ansatze. 
 To be specific, we compare the default ansatz in Eq.10 with the following  two fitting forms:

        \begin{align}   \label{eq:extrap_form2}
       xg(x,P_z,a)&=xg_0(x)+af_1(x)   \nonumber \\ & +\frac{d(x)}{P_z^2}+k(x)(m_{\pi}^2-m_{\pi,phy}^2),
       \end{align}
   \begin{align}  \label{eq:extrap_form3}
     xg(x,P_z,a)&=xg_0(x)+a^2f(x)   \nonumber \\ & +a^2 P_z^2h(x)+\frac{d(x)}{P_z^2}+k(x)(m_{\pi}^2-m_{\pi,phy}^2)
     .
     \end{align}

 We employ AIC-based model averaging to determine the combined central value and statistical uncertainty, and to estimate the systematic uncertainty associated with the choice of extrapolation ansatz.  The procedure is described below.

For the $i$th fit model, let $\chi_i^2$, $n_{i,\mathrm{par}}$, and $n_{i,\mathrm{data}}$ denote the total chi-square, the number of fit parameters, and the number of data points, respectively.  The normalized AIC weight is defined as 
\begin{align}
\omega_i &= \frac{\exp\left[-\frac{1}{2}(\chi_i^2+2n_{i,\mathrm{par}}-n_{i,\mathrm{data}})\right]}{\displaystyle\sum_j\exp\left[-\frac{1}{2}(\chi_j^2+2n_{j,\mathrm{par}}-n_{j,\mathrm{data}})\right]},
\end{align}

Suppose that  the $i$th  models  yields a central  value $m_i$ with  statistical uncertainty $\sigma_i$. We construct the cumulative distribution function (CDF) by integrating over  a weighted mixture of Gaussian distributions:
\begin{align}
P(y;\lambda) &= \int_{-\infty}^{y}dx\,
\sum_i\omega_i\frac{1}{\sigma_i\sqrt{\lambda}\sqrt{2\pi}}
\exp\!\left[-\frac{1}{2}\left(\frac{x-m_i}{\sigma_i\sqrt{\lambda}}\right)^2\right],
\end{align}
where $\lambda$ is introduced to control the width of the distribution for extracting statistical and  systematic uncertainties.

The quantiles $y_q(\lambda)$ is defined by
\begin{align}
P(y_q(\lambda);\lambda) &= q.
\end{align}
The  AIC-combined central value  can then be  estimated by the midpoint of the $16\%$ and $84\%$ quantiles at $\lambda=1$,
\begin{align}
m_{\mathrm{AIC}} &= \frac{y_{0.84}(1)+y_{0.16}(1)}{2},
\end{align}
and the  corresponding total uncertainty is defined as 
\begin{align}
\sigma_{\mathrm{tot}}(\lambda) &= \frac{y_{0.84}(\lambda)-y_{0.16}(\lambda)}{2}.
\end{align}

Assuming that the statistical and systematic uncertainties are independent, their variances satisfy
\begin{align}
\sigma_{\mathrm{tot}}^2(\lambda)
&=
\lambda\sigma_{\mathrm{stat}}^2
+
\sigma_{\mathrm{sys}}^2.
\end{align}
By repeating the analysis at $\lambda=1$ and $\lambda=2$, we can  obtain $\sigma_{\mathrm{stat}}$ and $\sigma_{\mathrm{sys}}$ by solving the resulting pair of equations.

\begin{figure*}    
     \centering
     \includegraphics[width=0.3\linewidth]{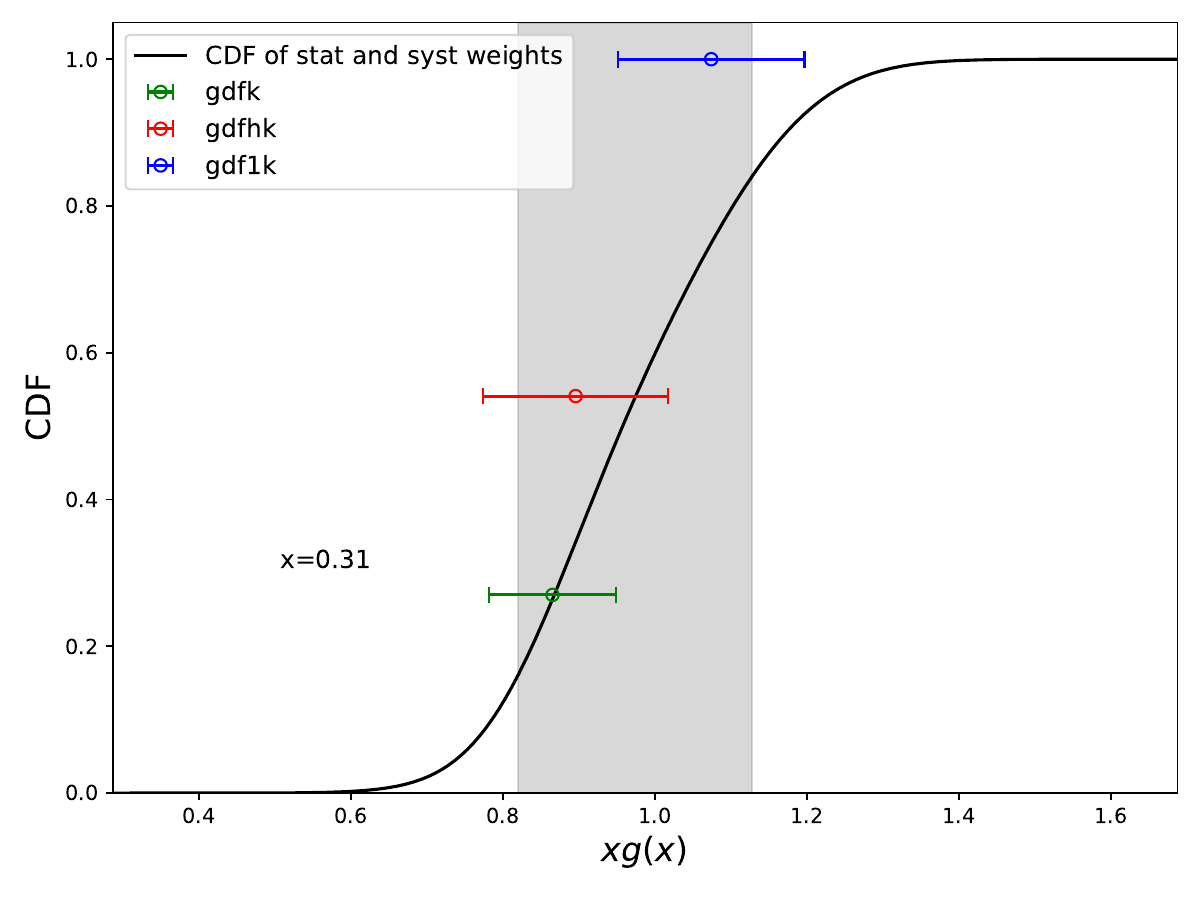}
      \includegraphics[width=0.3\linewidth]{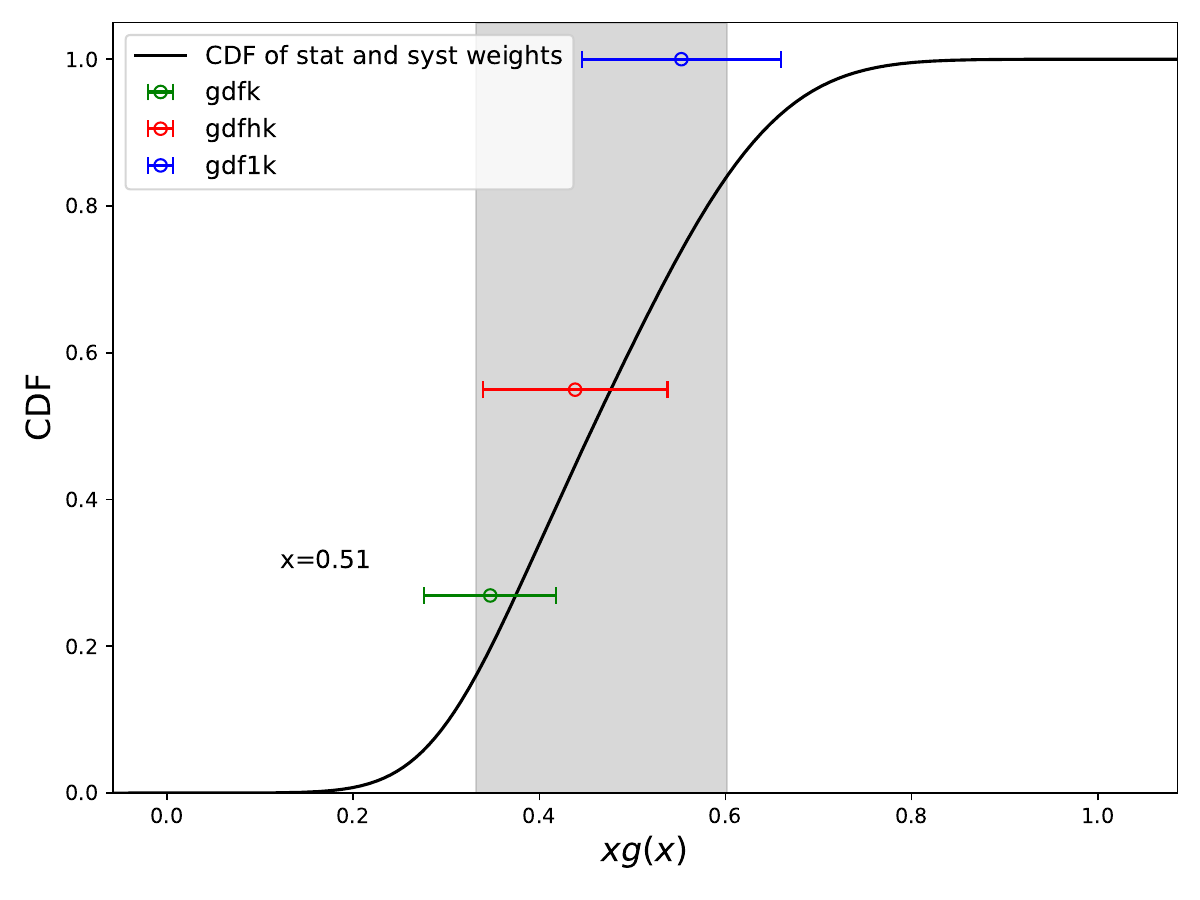}
       \includegraphics[width=0.3\linewidth]{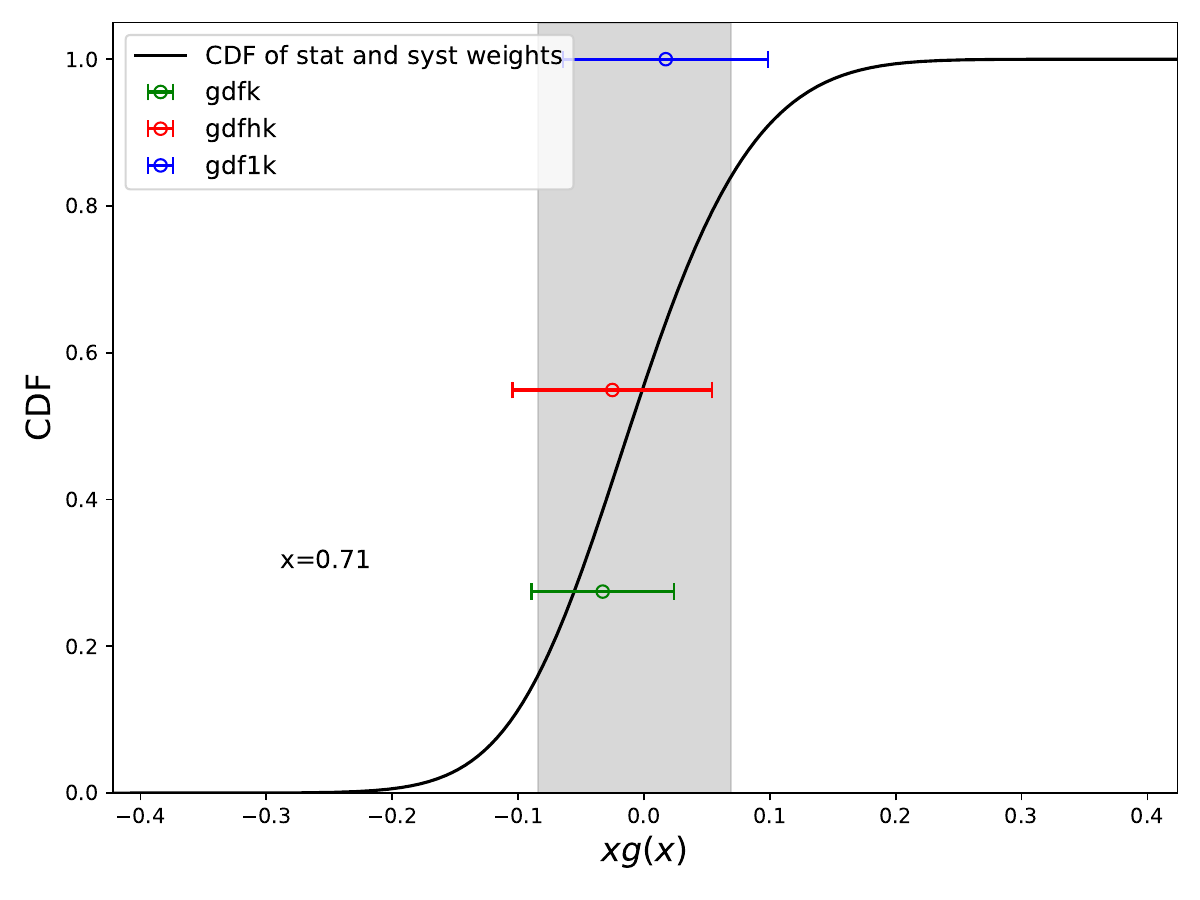}
  \caption{ The CDF for several representative values of   $x$.  The black curve denotes the combined CDF. The green, red and blue horizontal error bars show the extrapolated central values and statistical uncertainties obtained using Eq.10, Eq.\ref{eq:extrap_form3}, and Eq.\ref{eq:extrap_form2} respectively. The vertical position of each marker represents the cumulative AIC weight.  The gray band represents the central $68\%$ interval associated with the total uncertainty $\sigma_{\mathrm{tot}}(\lambda=1)$.}   \label{fig:AIC}
 \end{figure*}

\begin{figure*}
    \centering 
\includegraphics[width=1.0\linewidth]{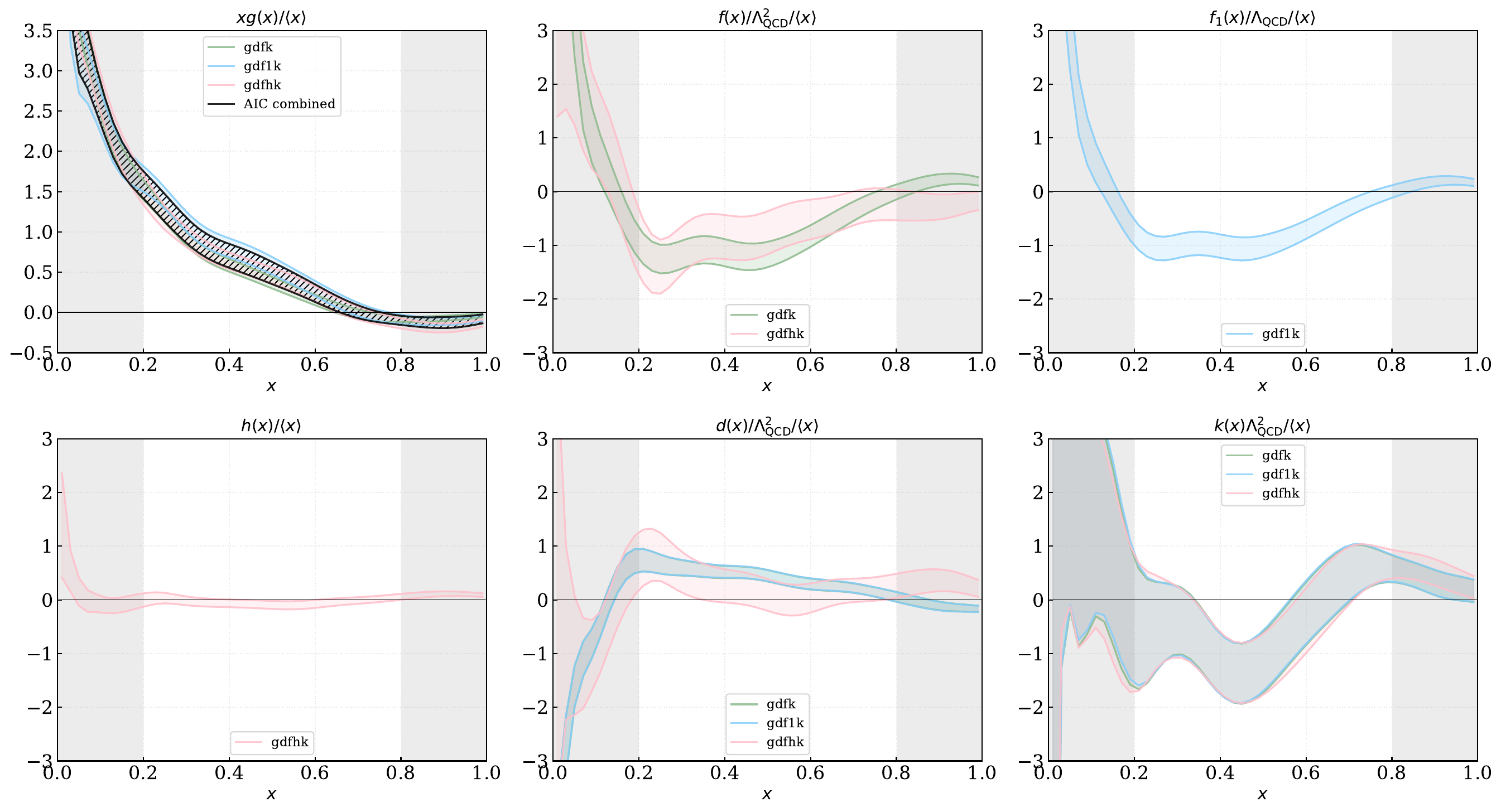}
    \caption{Contribution of different fitting terms of combined extrapolation. These parameters are combined with appropriate power of $\Lambda_\mathrm{QCD}$ to form dimensionless quantities, where $\Lambda_\mathrm{QCD}$  is taken to be $1$ GeV. We compare the results of different fitting formula including ~Eq.10(green band ),Eq.~\ref{eq:extrap_form2}(blue band) and Eq.~\ref{eq:extrap_form3}(red band). The AIC average of the extrapolated PDF is  represented as the black hatched band in the first subplot.}
    \label{fig:extrap_terms}
\end{figure*}

Fig.~\ref{fig:AIC} represents the CDF for several representative values of   $x$.  The black curve denotes the combined CDF. The colored horizontal error bars show the extrapolated central values and statistical uncertainties obtained using the different fitting  forms, ordered according to their central values. The vertical position of each marker represents the cumulative AIC weight.  As shown in the different panels, the extrapolated results obtained from the three fitting forms are mutually consistent within  uncertainties.  
The gray band spans the interval from $y_{0.16}(1)$ to $y_{0.84}(1)$ and therefore represents the central $68\%$ interval associated with the total uncertainty $\sigma_{\mathrm{tot}}(\lambda=1)$. The first panel of Fig.~\ref{fig:extrap_terms} shows the extrapolated results obtained from the different fitting forms at each value of $x$, together with the AIC-combined central value and its uncertainty band.

In Fig.~\ref{fig:extrap_terms} we also represent the fitting results of each parameter including $f(x)$, $h(x)$ and $d(x)$. These parameters are combined with appropriate powers of $\Lambda_\mathrm{QCD}$ to form dimensionless quantities, where $\Lambda_\mathrm{QCD}$  is taken to be $1$ GeV. As can be seen from the plots, except for $h(x)$ parameter, which value is equal to zero within uncertainties, all other dimensionless quantities are of order $O(1)$, which means that these terms are of $O(\Lambda_\mathrm{QCD})$ size when held fixed in physics unit.

According to the LaMET power-counting argument ~\cite{Braun:2018brg, Liu:2020rqi, Ji:2022ezo}, the higher-order corrections are suppressed by the scales $x^2P_z^2$ and $(1-x)^2P_z^2$. Therefore, the LaMET expansion is expected to be reliable mainly in the intermediate-$x$ region, while it breaks down toward the endpoint regions $x\to 0$ and $x\to 1$, where corrections of the form $\Lambda_{\rm QCD}^2/(x^2P_z^2)$ and $\Lambda_{\rm QCD}^2/[(1-x)^2P_z^2]$ can become large. However we do not observe a clear $1/x^2$ or $1/(1-x)^2$ structure of fitted $d(x)$ in the figure, since at these endpoint regions LaMET expansion itself is no longer under perturbative control. 

\subsection{Estimation of  systematic and statistic uncertainties}

\begin{figure}
    \centering 
\includegraphics[width=1.0\linewidth]{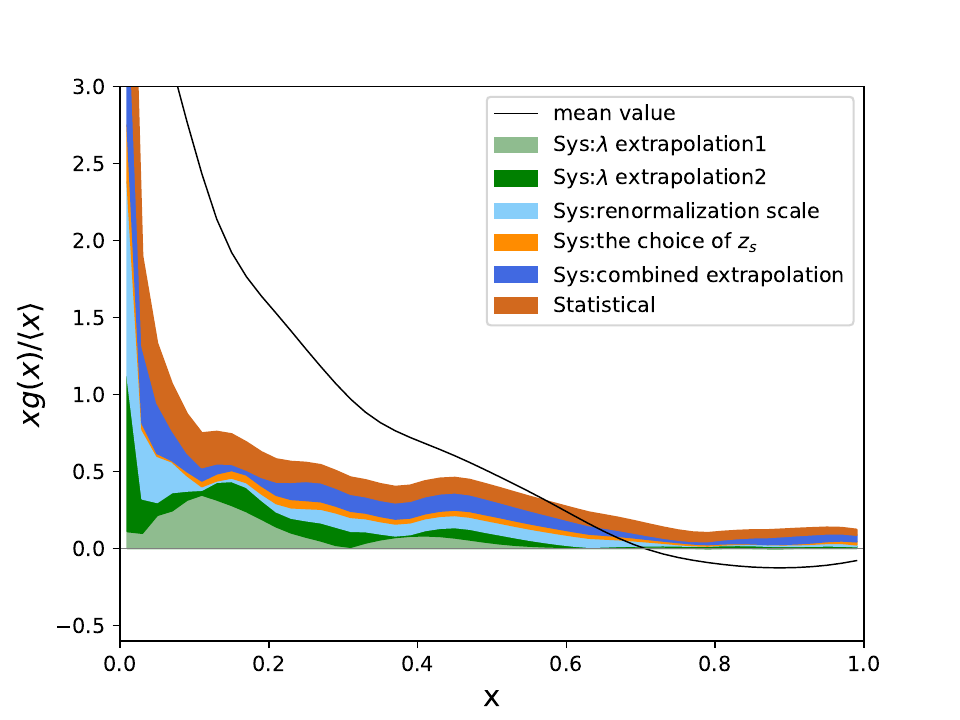}
    \caption{Estimation of statistical and systematic uncertainties  in combined extrapolation. The width of the (nonoverlapping) colored bands denotes the size of each uncertainty (these uncertainties are
    added in quadrature and the square root is taken to obtain
    the full uncertainty). The black curve is the mean value
    obtained at $z_s = 0.3$ fm, $\mu=2$ GeV, the default lambda extrapolation range shown in Fig.~\ref{fig:lambda_extrap_part1}, and  extrapolated to continuum limit, infinite momentum limit and physical pion mass.}
     \label{fig:sta_sys_error}
\end{figure}

In Fig.~\ref{fig:sta_sys_error} we show the central value of the gluon PDF together with  the contributions of all  uncertainties sources. The systematic uncertainties of our final result arise from the $\lambda$ extrapolation, the renormalization scale dependence, the choice of $z_s$ and  the combined extrapolation.

\textit{$\lambda$ extrapolation:} In the default analysis, we use the fitting form in Eq.~\ref{eq:lambda_extrap}, and the corresponding fitting ranges are indicated by the gray bands in Fig.~\ref{fig:lambda_extrap_part1}. The uncertainty associated with the $\lambda$ extrapolation mainly comes from two sources.

The first source is the choice of the fitting range. To estimate this uncertainty, we reduce the upper limit of the fitting range by one lattice spacing. The difference between the central values obtained from the default and modified fitting ranges is taken as the corresponding systematic uncertainty, which is shown as the light green band.

The second source of uncertainties is  associated with the choice of the parametrization form. To evaluate it, we also tested the following form proposed in Ref.~\cite{Ji:2026vir},
 \begin{align}
	  \tilde{h}_R(\lambda) &= (l_1 \lambda + l_2) e^{-\lambda/\lambda_0}.
	  \label{eq:lambda_extrap_NLA}
\end{align} 
The difference between the results obtained using this form and our default parametrization is taken as an additional contribution to the systematic uncertainty.


\textit{Renormalization scale dependence:} In the default analysis, we set the renormalization scale to $\mu=2$ GeV, corresponding to $\alpha_s(\mu)=0.296$. To estimate the systematic uncertainty associated with the renormalization scale dependence, we repeat the analysis using $\mu=4$ GeV and $\alpha_s(\mu)=0.23$. The difference between the central values obtained at the two scales is taken as the corresponding systematic uncertainty, which is shown as the light-blue band.

\textit{Choice of $z_s$ in the hybrid scheme:} The systematic uncertainty associated with the choice of $z_s$, shown as the orange band, is estimated by taking the difference between the results obtained with $z_s=0.3$ fm, used in the default analysis, and those obtained with $z_s=0.2$ fm.

 \textit{Combined extrapolation}: The systematic uncertainty associated with the combined extrapolation
is estimated by doing the AIC analysis using different extrapolation forms, as have been mentioned above.
The systematic uncertainty brought by combined extrapolation is shown as the dark blue band.


All the systematic uncertainties, together with statistical uncertainty
are added in quadrature and  the square root is taken to obtain the full  uncertainty.

\vspace{10em}

\clearpage


\end{document}